%% file: ArXiv_1609_06512_v2.tex
\RequirePackage{amssymb}
\RequirePackage{amsfonts}
\RequirePackage{amsmath}
\documentclass[10pt]{iopart}
\usepackage{euscript}
\usepackage{frcursive}
\usepackage{color}
\usepackage{graphicx}
\usepackage{rotating}
\usepackage{bm}
\usepackage{curves}
\usepackage[svgnames,dvipsnames]{pstricks}
\usepackage{lettrine}
\usepackage{mathrsfs}
\usepackage{booktabs}
\usepackage{scalefnt}
\usepackage[normalem]{ulem}
\usepackage{dcolumn}
\usepackage{tikz}
\DeclareGraphicsExtensions{{},.pdf}
\usepackage[dcucite]{harvard}
\providecommand{\tsn}[1]{{\textup{\scalefont{0.80}#1}}}

\providecommand{\tabref}[1]{{\textup{(Tab.~\ref{#1})}}}
\providecommand{\figref}[1]{{\textup{(Fig.~\ref{#1})}}}

\providecommand{\tabrefnp}[1]{{\textup{Tab.~\ref{#1}}}}
\providecommand{\figrefnp}[1]{{\textup{Fig.~\ref{#1}}}}

\providecommand{\eqrefsatob} [2]{\textup{(\ref{#1}--\ref{#2})}}
\providecommand{\eqrefsab}   [2]{\textup{(\ref{#1}, \ref{#2})}}
\providecommand{\eqrefsabc}  [3]{\textup{(\ref{#1}, \ref{#2}, \ref{#3})}}
\providecommand{\eqrefsabcd} [4]{\textup{(\ref{#1}, \ref{#2}, \ref{#3}, \ref{#4})}}

\providecommand{\tabrefsab}   [2]{{\textup{(Tabs.~\ref{#1}, \ref{#2})}}}
\providecommand{\figrefsab}   [2]{{\textup{(Figs.~\ref{#1}, \ref{#2})}}}

\providecommand{\parref}[1]{{\textup{(\S\ref{#1})}}}

\providecommand{\parrefnp}[1]{{\textup{\S\ref{#1}}}}

\providecommand{\const}{{\rm const}}

\providecommand{\ie}{{\em ie}}
\providecommand{\eg}{{\em eg}}
\providecommand{\viz}{{\em viz}}

\providecommand{\TKE}{{\rm k}}
\newcommand{\tsr}[1]{{\boldsymbol{\mathbf{#1}}}}

\newcommand{\II}[1]{ \mathrm{II}_{\boldsymbol{\mathbf{#1}}}}
\newcommand{\III}[1]{\mathrm{III}_{\boldsymbol{\mathbf{#1}}}}

\makeatletter
\newcommand{\vast}{\bBigg@{4}}
\newcommand{\Vast}{\bBigg@{5}}
\makeatother
\def\NumERICCS{N\kern-.09em\lower.5ex\hbox{\tsn{um}}\kern-.09em\tsn{ERICCS}}
\begin{document}
%
%
%
%
%
%
%
%
%
\title[$\varepsilon_{ij}$-budgets in turbulent plane channel flow]{Further analysis of the budgets of the dissipation tensor $\varepsilon_{ij}$ in turbulent plane channel flow}
%
%
%
%
%
%
%
%
%
\author{G. A. Gerolymos and I. Vallet}
\address{Sorbonne Universit\'es, Universit\'e Pierre-et-Marie-Curie, 4 place Jussieu, 75005 Paris, France}
\ead{{\color{blue}{georges.gerolymos@upmc.fr}} and {\color{blue}{isabelle.vallet@upmc.fr}}}
\date{\today}
\begin{abstract}
Recent \tsn{DNS} results [Gerolymos G.A., Vallet I. : {\it J. Fluid Mech.} {\bf 807} (2016) 386--418] have provided data for the terms in the transport equations for the components
of the dissipation tensor $\varepsilon_{ij}$ in low-Reynolds turbulent plane channel flow. The present paper extends the previous results by a detailed analysis
of the behaviour of various mechanisms in the $\varepsilon_{ij}$-transport equations (production, diffusion, redistribution, destruction),
with particular emphasis on the component-by-component comparison with the corresponding mechanisms in the transport equations for the Reynolds-stresses $r_{ij}$.
The splitting of the pressure terms for the wall-normal components into redistribution and pressure-diffusion reveals substantially different behaviour near the wall.
The wall-asymptotics of different terms in the transport equations are studied in detail, and examined using the \tsn{DNS} data.
Both \tsn{DNS} data and wall-asympotic analysis show that the anisotropy of the destruction-of-dissipation tensor $\varepsilon_{\varepsilon_{ij}}$ is fundamentally different from that of $r_{ij}$ or $\varepsilon_{ij}$,
never approaching the 2-component (2-C) state at the solid wall.
\end{abstract}
%
%
%
%
%
%
%
%
%
%
\section{Introduction}\label{FAepsijBsTPCF_s_I}
%
%
%
%
%
%
%
%
%

Transport equations \cite{Chou_1945a} of 1-point and 2-point statistics are essential both in understanding turbulence dynamics \cite{Tennekes_Lumley_1972a}
and in providing the theoretical foundations for turbulence modelling \cite{Schiestel_2008a}. The fluctuating-velocity-covariance (2-moment) tensor $r_{ij}:=\overline{u_i'u_j'}$, which defines the Reynolds-stresses $-\rho r_{ij}$,
is governed by well known transport equations \cite[(1), p. 17]{Mansour_Kim_Moin_1988a} where the dissipation tensor $\varepsilon_{ij}$ represents the destruction of $r_{ij}$ by molecular friction (viscosity).
The dissipation tensor $\varepsilon_{ij}$ also follows transport equations \cite[(3.3), p. 403]{Gerolymos_Vallet_2016b} where the destruction-of-dissipation tensor $\varepsilon_{\varepsilon_{ij}}$
represents the destruction of $\varepsilon_{ij}$ by molecular viscosity. Of course $\varepsilon_{\varepsilon_{ij}}$ is governed in turn by its own transport equation where appears its own destruction-rate, and so on to correlations
of higher derivatives of the fluctuating velocity.

The budgets of the $r_{ij}$-transport equations \eqref{Eq_FAepsijBsTPCF_s_TEqsWAs_ss_TEqs_001a} have been studied extensively using \tsn{DNS} \cite{Mansour_Kim_Moin_1988a,Moser_Kim_Mansour_1999a,Sillero_Jimenez_Moser_2013a}.
Closure of noncomputable terms in \eqref{Eq_FAepsijBsTPCF_s_TEqsWAs_ss_TEqs_001a}, along with a transport equation for some scalar
scale-determining variable \cite{Jones_Launder_1972a,Launder_Spalding_1974a,Wilcox_1988a,Menter_1994a,Jakirlic_Hanjalic_2002a}
has led \cite{Launder_Reece_Rodi_1975a} to the development of second-moment closures (\tsn{SMC}s) or Reynolds-stress models (\tsn{RSM}s).
Several models of this family have been assessed for the computation of complex 3-D flows \cite{Gerolymos_Vallet_2001a,Jakirlic_Eisfeld_JesterZurker_Kroll_2007a,Cecora_Radespiel_Eisfeld_Probst_2015a}
and are increasingly used to predict practical 3-D configurations \cite{Eisfeld_ed_2015a}.
Comparisons with measurements \cite{Rumsey_NASATurbmodels} demonstrate the predictive improvement of 7-equation \tsn{RSM}s against standard 2-equation approaches,
especially in presence of separation and/or secondary flows \cite{Gerolymos_Joly_Mallet_Vallet_2010a,Gerolymos_Vallet_2016a}
but also highlight remaining challenges.
In general \tsn{RSM}s cannot return the correct wall-asymptotic behaviour for all of the components of the Reynolds-stress tensor \cite{Yakovenko_Chang_2007a},
and privileging the wall-normal components improves log-law prediction \cite{Gerolymos_Lo_Vallet_Younis_2012a}. An even more difficult challenge is to
correctly mimic the $Re$-dependence of the near-wall maxima of the diagonal Reynolds-stresses which is revealed by \tsn{DNS} results \cite{Lee_Moser_2015a}.
Finally, the hysteretic behaviour of the separation-and-reattachment process \cite{Gerolymos_Kallas_Papailiou_1989a} may require additional specific lag-treatments \cite{Olsen_Coakley_2001a}.
                                                                               
The correct prediction of near-wall anisotropy \cite{Durbin_1993a} and of lengthscale anisotropy in general \cite{Lumley_Yang_Shih_1999a} is necessary to meet these challenges.
The replacement of the scalar scale-determining equation used in classical \tsn{RSM}s \cite{Wilcox_2006a,Schiestel_2008a}
by transport equations for the individual components of $\varepsilon_{ij}$
has been suggested to overcome the unsatisfactory {\em a posteriori} perfomance of algebraic $\varepsilon_{ij}$-closures \cite{Gerolymos_Lo_Vallet_Younis_2012a}.
Detailed \tsn{DNS} data of the $\varepsilon_{ij}$-transport equations \eqref{Eq_FAepsijBsTPCF_s_TEqsWAs_ss_TEqs_001b} are necessary to achieve this goal.

Scrutiny of the budgets of the scalar $\varepsilon$-equation ($\varepsilon:=\tfrac{1}{2}\varepsilon_{mm}$) provided by \tsn{DNS} \cite{Mansour_Kim_Moin_1988a}
has proved particularly useful in improving the closure of this equation \cite{Lai_So_1990a,Rodi_Mansour_1993a,Jakirlic_Hanjalic_2002a}.
On the other hand, very little work has been done concerning the budgets of the tensorial $\varepsilon_{ij}$-equations \eqref{Eq_FAepsijBsTPCF_s_TEqsWAs_ss_TEqs_001b}.
In a recent work \cite{Gerolymos_Vallet_2016b} we have generated \tsn{DNS} data of $\varepsilon_{ij}$-budgets for low-$Re$ turbulent plane channel flow
and discussed the behaviour of various terms in \eqref{Eq_FAepsijBsTPCF_s_TEqsWAs_ss_TEqs_001b}, with particular emphasis on the 4 production mechanisms.

The purpose of the present work is to further analyze $\varepsilon_{ij}$-budgets in turbulent plane channel flow, and in particular the similarities and differences with respect to $r_{ij}$-budgets.
In \parrefnp{FAepsijBsTPCF_s_TEqsWAs} we define the terms in the transport equations for $r_{ij}$ and $\varepsilon_{ij}$,
and calculate the wall-asymptotic behaviour of different terms in the $\varepsilon_{ij}$-transport
equations \eqref{Eq_FAepsijBsTPCF_s_TEqsWAs_ss_TEqs_001b} for the particular case of turbulent plane channel flow.
These analytical results are used \parref{FAepsijBsTPCF_s_TPCFBs} to assess very-near-wall \tsn{DNS} data.
In \parrefnp{FAepsijBsTPCF_s_TPCFBs} we use \tsn{DNS} data \parref{FAepsijBsTPCF_s_TPCFBs_ss_DNSCs} to compare
$r_{ij}$-budgets with $\varepsilon_{ij}$-budgets \parref{FAepsijBsTPCF_s_TPCFBs_ss_epsijvsrijBs} and to analyze the splitting of the pressure term $\Pi_{\varepsilon_{ij}}$ in \eqref{Eq_FAepsijBsTPCF_s_TEqsWAs_ss_TEqs_001b}
into a redistributive and a conservative term \parref{FAepsijBsTPCF_s_TPCFBs_ss_RPD}.
In \parrefnp{FAepsijBsTPCF_s_epsepsij} we compare the anisotropy and associated anisotropy invariant mapping (\tsn{AIM})
of the Reynolds-stresses $r_{ij}$, their dissipation $\varepsilon_{ij}$ and the destruction-of-dissipation $\varepsilon_{\varepsilon_{ij}}$ which exhibits a notably different componentality near the wall.
Finally, in \parrefnp{FAepsijBsTPCF_s_C}, we summarize the main results of the present work.

%
%
%
%
%
%
%
%
%
\section{Transport equations and wall asymptotics}\label{FAepsijBsTPCF_s_TEqsWAs}
%
%
%
%
%
%
%
%
%

Consistent with the \tsn{DNS} data, we study incompressible flow with a Newtonian constitutive relation in an inertial frame \cite{Gerolymos_Vallet_2016b}.
We use a Cartesian reference-frame $x_i\in\{x,y,z\}$, note $u_i\in\{u,v,w\}$ the corresponding components of the velocity vector,
and use Reynolds decomposition into averaged $\overline{(\cdot)}$ and fluctuating $(\cdot)'$ quantities, We note $t$ the time, $\rho\approxeq\const$ the density,
$p$ the pressure, $\nu\approxeq\const$ the kinematic viscosity, and $\mu:=\rho\nu\approxeq\const$ the dynamic viscosity.

%
%
%
%
%
\subsection{Transport equations}\label{FAepsijBsTPCF_s_TEqsWAs_ss_TEqs}
%
%
%
%
%

Straightforward manipulations of the fluctuating momentum \eqref{Eq_FAepsijBsTPCF_s_AppendixABVSy+0_ss_PCF_sss_WNFMFPF_001}
and of the fluctuating continuity \eqref{Eq_FAepsijBsTPCF_s_AppendixABVSy+0_ss_FCEq_002} equations and of their gradients
lead to the transport equations for $r_{ij}:=\overline{u_i'u_j'}$ \cite[(1), p. 17]{Mansour_Kim_Moin_1988a}
\begin{subequations}
                                                                                                                                    \label{Eq_FAepsijBsTPCF_s_TEqsWAs_ss_TEqs_001}
\begin{alignat}{6}
\underbrace{\rho\dfrac{\partial\overline{u'_i u_j'}}
                      {\partial                  t }+\rho\bar u_\ell\dfrac{\partial\overline{u'_i u_j'}}
                                                                          {\partial              x_\ell}}_{\displaystyle C_{ij}}=&\underbrace{\dfrac{\partial       }
                                                                                                                                                    {\partial x_\ell}\Bigg(\mu\dfrac{\partial\overline{u'_iu_j'}}
                                                                                                                                                                                    {\partial x_\ell            }\Bigg)}_{\displaystyle d_{ij}^{(\mu)}}
                                                                                                                                + \underbrace{\dfrac{\partial       }
                                                                                                                                                    {\partial x_\ell}\Big(-\rho\overline{u'_iu_j'u'_\ell}\Big)}_{\displaystyle d_{ij}^{(u)}}
                                                                                                                                + \underbrace{\Bigg(-\overline{u_i'\dfrac{\partial p' }
                                                                                                                                                                         {\partial x_j}}
                                                                                                                                                    -\overline{u_j'\dfrac{\partial p' }
                                                                                                                                                                         {\partial x_i}}\Bigg)}_{\displaystyle \Pi_{ij}}
                                                                                                                                    \notag\\
                                                                                                                                +&\underbrace{\Bigg(-\rho\overline{u_i'u_\ell'}\dfrac{\partial\bar u_j}
                                                                                                                                                                                     {\partial  x_\ell}
                                                                                                                                                    -\rho\overline{u_j'u_\ell'}\dfrac{\partial\bar u_i}
                                                                                                                                                                                     {\partial  x_\ell}\Bigg)}_{\displaystyle P_{ij}}
                                                                                                                                - \underbrace{\Bigg(2\mu\overline{\dfrac{\partial u_i'  }
                                                                                                                                                                        {\partial x_\ell}
                                                                                                                                                                  \dfrac{\partial u_j'  }
                                                                                                                                                                        {\partial x_\ell}}\Bigg)}_{\displaystyle\rho\varepsilon_{ij}}
                                                                                                                                    \label{Eq_FAepsijBsTPCF_s_TEqsWAs_ss_TEqs_001a}
\end{alignat}
and $\varepsilon_{ij}$ \cite[(3.3), p. 403]{Gerolymos_Vallet_2016b}
\begin{alignat}{6}
&\underbrace{\rho\dfrac{\partial\varepsilon_{ij}}
                       {\partial               t}+\rho\bar u_\ell\dfrac{\partial\varepsilon_{ij}}
                                                                       {\partial          x_\ell}}_{\displaystyle C_{\varepsilon_{ij}}}=
  \underbrace{\dfrac{\partial       }
                    {\partial x_\ell}\left[\mu\dfrac{\partial\varepsilon_{ij}}
                                                    {\partial x_\ell          }\right]}_{\displaystyle d_{\varepsilon_{ij}}^{(\mu)}}
 +\underbrace{\dfrac{\partial       }
                    {\partial x_\ell}\left[-\rho\left(\overline{u'_\ell2\nu\dfrac{\partial u'_i}
                                                                                 {\partial  x_k}\dfrac{\partial u'_j}
                                                                                                      {\partial  x_k}}\right)\right]}_{\displaystyle d_{\varepsilon_{ij}}^{(u)}}
                                                                                                                                    \notag\\
&\underbrace{-\rho\varepsilon_{i\ell}\dfrac{\partial\bar u_j}
                                           {\partial  x_\ell}
             -\rho\varepsilon_{j\ell}\dfrac{\partial\bar u_i}
                                           {\partial  x_\ell}}_{\displaystyle P_{\varepsilon_{ij}}^{(1)}}
 \underbrace{-\rho\left(2\nu\overline{\dfrac{\partial u'_i}
                                            {\partial  x_k}\dfrac{\partial   u'_j}
                                                                 {\partial x_\ell}}\right)\left(\dfrac{\partial\bar u_k   }
                                                                                                      {\partial     x_\ell}+\dfrac{\partial\bar u_\ell}
                                                                                                                                  {\partial     x_k   }\right)}_{\displaystyle P_{\varepsilon_{ij}}^{(2)}}
                                                                                                                                    \notag\\
&\underbrace{-\rho\left(2\nu\overline{u'_\ell\dfrac{\partial u'_i}
                                                   {\partial  x_k}}\right)\dfrac{\partial^2 \bar u_j}
                                                                                {\partial x_\ell \partial x_k}
             -\rho\left(2\nu\overline{u'_\ell\dfrac{\partial u'_j}
                                                   {\partial  x_k}}\right)\dfrac{\partial^2 \bar u_i}
                                                                                {\partial x_\ell \partial x_k}}_{\displaystyle P_{\varepsilon_{ij}}^{(3)}}
 \underbrace{-\rho\left[2\nu\overline{\dfrac{\partial u'_\ell}
                                            {\partial     x_k}\left(\dfrac{\partial u'_i}
                                                                          {\partial  x_k}\dfrac{\partial   u'_j}
                                                                                               {\partial x_\ell}
                                                                   +\dfrac{\partial u'_j}
                                                                          {\partial  x_k}\dfrac{\partial   u'_i}
                                                                                               {\partial x_\ell}\right)}\right]}_{\displaystyle \Xi_{\varepsilon_{ij}}=:P_{\varepsilon_{ij}}^{(4)}}
                                                                                                                                    \notag\\
&\underbrace{-2\nu\overline{\dfrac{\partial u'_i}
                                  {\partial  x_k}\dfrac{\partial^2            p'}
                                                       {\partial x_j\partial x_k}}
             -2\nu\overline{\dfrac{\partial u'_j}
                                  {\partial  x_k}\dfrac{\partial^2            p'}
                                                       {\partial x_i\partial x_k}}}_{\displaystyle \Pi_{\varepsilon_{ij}}}
 -\underbrace{\rho\overline{\left(2\nu\dfrac{\partial^2 u'_i}
                                            {\partial  x_k\partial x_\ell}\right)\left(2\nu\dfrac{\partial^2 u'_j}
                                                                                                 {\partial  x_k\partial x_\ell}\right)}}_{\displaystyle \rho\varepsilon_{\varepsilon_{ij}}}
                                                                                                                                    \label{Eq_FAepsijBsTPCF_s_TEqsWAs_ss_TEqs_001b}
\end{alignat}
\end{subequations}
which were reproduced here for completness.

The common origin of \eqrefsab{Eq_FAepsijBsTPCF_s_TEqsWAs_ss_TEqs_001a}
                              {Eq_FAepsijBsTPCF_s_TEqsWAs_ss_TEqs_001b}
leads to analogous mechanisms in both transport equations, where convection by the mean flow ($C_{ij}, C_{\varepsilon_{\ij}}$)
is balanced by 5 mechanisms: diffusion by molecular viscosity ($d_{ij}^{(\mu)}, d_{\varepsilon_{ij}}^{(\mu)}$),
turbulent diffusion (mixing) by the fluctuating velocity field $u_\ell'$ ($d_{ij}^{(u)}, d_{\varepsilon_{ij}}^{(u)}$),
production by various mechanisms ($P_{ij}, P_{\varepsilon_{ij}}:=P_{\varepsilon_{ij}}^{(1)}+P_{\varepsilon_{ij}}^{(2)}+P_{\varepsilon_{ij}}^{(3)}+P_{\varepsilon_{ij}}^{(4)}$),
the fluctuating-pressure mechanisms ($\Pi_{ij}, \Pi_{\varepsilon_{ij}}$), and destruction by molecular viscosity ($\varepsilon_{\ij}, \varepsilon_{\varepsilon_{ij}}$).
Of course the tensorial componentality \cite{Lumley_1978a,Kassinos_Reynolds_Rogers_2001a,Simonsen_Krogstad_2005a}
and the scaling \cite[pp. 88--92]{Tennekes_Lumley_1972a} of various terms in \eqref{Eq_FAepsijBsTPCF_s_TEqsWAs_ss_TEqs_001b} differs from that of the corresponding terms in \eqref{Eq_FAepsijBsTPCF_s_TEqsWAs_ss_TEqs_001a}.

%
%
%
%
%
%
%
%
%
\subsection{Wall asymptotics}\label{FAepsijBsTPCF_s_TEqsWAs_ss_WAs}
%
%
%
%
%
%
%
%
%

Before studying the present \tsn{DNS} data for the $\varepsilon_{ij}$-transport budgets \eqref{FAepsijBsTPCF_s_TPCFBs_ss_epsijvsrijBs}, it is useful to summarize the theoretically expected (\ref{FAepsijBsTPCF_s_AppendixABVSy+0})
asymptotic behaviour of various terms in the viscous sublayer, or, formally, as $y^+\to0$. Inner scaling \cite[$\cdot^+$]{Buschmann_GadelHak_2007a} is consistently used in these calculations (\ref{FAepsijBsTPCF_s_AppendixABVSy+0}).
Wall-asymptotics of the terms in \eqref{Eq_FAepsijBsTPCF_s_TEqsWAs_ss_TEqs_001b} which only involve fluctuating velocities and their derivatives
($d_{\varepsilon_{ij}}^{(\mu)}$, $d_{\varepsilon_{ij}}^{(u)}$, $P_{\varepsilon_{ij}}^{(4)}$, $\varepsilon_{\varepsilon_{ij}}$) can be readily obtained from the
Taylor-series expansions \cite[\S4.6, pp. 136--141]{Riley_Hobson_Bence_2006a} of $u_i'^+$ in the wall-normal direction $y^+$
\begin{alignat}{6}
(\cdot)'^+\underset{y^+\to0}{\sim}(\cdot)_w'^+(x^+,z^+,t^+)&+&A_{(\cdot)}'^+(x^+,z^+,t^+)\;y^+    &+&B_{(\cdot)}'^+(x^+,z^+,t^+)\;{y^+}^2& &
                                                                                                                                    \notag\\
                                                           &+&C_{(\cdot)}'^+(x^+,z^+,t^+)\;{y^+}^3&+&D_{(\cdot)}'^+(x^+,z^+,t^+)\;{y^+}^4&+&\cdots
                                                                                                                                    \label{Eq_FAepsijBsTPCF_s_WAs_001}
\end{alignat}
under the constraints of the no-slip condition at the wall \eqref{Eq_FAepsijBsTPCF_s_AppendixFDPCF_ss_MFS_001a} and of the fluctuating continuity equation \parref{FAepsijBsTPCF_s_AppendixABVSy+0_ss_FCEq}.
On the contrary, determination of the wall-asymptotics of terms in \eqref{Eq_FAepsijBsTPCF_s_TEqsWAs_ss_TEqs_001b} which contain the fluctuating pressure and its derivatives ($\Pi_{\varepsilon_{ij}}$) or the
mean-flow velocities and their derivatives ($C_{\varepsilon_{\ij}}$, $P_{\varepsilon_{ij}}^{(1)}$, $P_{\varepsilon_{ij}}^{(2)}$, $P_{\varepsilon_{ij}}^{(3)}$),
requires specific simplifications implied by the fully developed plane channel flow conditions \eqrefsatob{Eq_FAepsijBsTPCF_s_AppendixFDPCF_ss_MFS_001}
                                                                                                         {Eq_FAepsijBsTPCF_s_AppendixFDPCF_ss_MFS_003},
in line with the analysis of the budgets of $r_{ij}$ and $\varepsilon$ in \citeasnoun{Mansour_Kim_Moin_1988a}.
Using \eqrefsabc{Eq_FAepsijBsTPCF_s_WAs_001}
                {Eq_FAepsijBsTPCF_s_AppendixABVSy+0_ss_FCEq_003}
                {Eq_FAepsijBsTPCF_s_AppendixABVSy+0_ss_FCEq_004},
along with specific results \eqrefsatob{Eq_FAepsijBsTPCF_s_AppendixABVSy+0_ss_PCF_sss_MF_001}
                                       {Eq_FAepsijBsTPCF_s_AppendixABVSy+0_ss_PCF_sss_WPFM_004}
applicable to plane channel flow satisfying conditions \eqrefsatob{Eq_FAepsijBsTPCF_s_AppendixFDPCF_ss_MFS_001}
                                                                  {Eq_FAepsijBsTPCF_s_AppendixFDPCF_ss_MFS_003},
readily yields the wall-asymptotic expansions \tabrefsab{Tab_FAepsijBsTPCF_s_WAs_001}
                                                        {Tab_FAepsijBsTPCF_s_WAs_002}
of various terms in the $\varepsilon_{ij}$-transport equations \eqref{Eq_FAepsijBsTPCF_s_TEqsWAs_ss_TEqs_001b}.
The homogeneity relations \eqref{Eq_FAepsijBsTPCF_s_AppendixFDPCF_ss_MFS_003} were used, when applicable to simplify these expressions.
The plane channel flow identity $\overline{B_u'^+C_v'^+}\stackrel{\eqref{Eq_FAepsijBsTPCF_s_AppendixABVSy+0_ss_PCF_sss_WPFM_003c}}{=}0$ was used in
$\varepsilon_{\varepsilon_{ij}}^+$, $d_{\varepsilon_{xy}}^{(\mu)+}$ and $\Pi_{\varepsilon_{xy}}^+$ \tabref{Tab_FAepsijBsTPCF_s_WAs_001},
while the plane channel flow identity \eqref{Eq_FAepsijBsTPCF_s_AppendixABVSy+0_ss_PCF_sss_WPFM_004} was used to replace $\overline{B_v'\partial_tA_u'}^+$ in $\Pi_{\varepsilon_{xy}}^+$ \tabref{Tab_FAepsijBsTPCF_s_WAs_001}.
These results \tabrefsab{Tab_FAepsijBsTPCF_s_WAs_001}
                        {Tab_FAepsijBsTPCF_s_WAs_002}
are used in the analysis of the \tsn{DNS} data \parref{FAepsijBsTPCF_s_TPCFBs}.
\begin{table}
\begin{center}
\rule{\textwidth}{.25pt}
\input{Tab_FluidDynRes_01_WAs_epsijT}
\caption{Asymptotic (as $y^+\to0$) expansions \eqref{Eq_FAepsijBsTPCF_s_WAs_001}
         of various terms ($d_{\varepsilon_{ij}}^{(\mu)}$, $d_{\varepsilon_{ij}}^{(u)}$, $d_{\varepsilon_{ij}}^{(p)}$, $\phi_{\varepsilon_{ij}}$, $\Pi_{\varepsilon_{ij}}$)
         in the $\varepsilon_{ij}$-transport equations \eqref{Eq_FAepsijBsTPCF_s_TEqsWAs_ss_TEqs_001b},
         in wall-units \cite[(A3), p. 414]{Gerolymos_Vallet_2016b}, for the particular case of plane channel flow \eqrefsatob{Eq_FAepsijBsTPCF_s_AppendixFDPCF_ss_MFS_001}
                                                                                                                                  {Eq_FAepsijBsTPCF_s_AppendixFDPCF_ss_MFS_003}.}
\label{Tab_FAepsijBsTPCF_s_WAs_001}
\rule{\textwidth}{.25pt}
\end{center}
\end{table}
%
\begin{table}
\begin{center}
\rule{\textwidth}{.25pt}
\input{Tab_FluidDynRes_01_WAs_epsepsij_Pepsij}
\caption{Asymptotic (as $y^+\to0$) expansions \eqref{Eq_FAepsijBsTPCF_s_WAs_001}
         of the various mechanisms of production $P_{\varepsilon_{ij}}=P^{(1)}_{\varepsilon_{ij}}+P^{(2)}_{\varepsilon_{ij}}+P^{(3)}_{\varepsilon_{ij}}+P^{(4)}_{\varepsilon_{ij}}$
         and of the destruction-of-dissipation $\varepsilon_{\varepsilon_{ij}}$
         appearing in the $\varepsilon_{ij}$-transport equations \eqref{Eq_FAepsijBsTPCF_s_TEqsWAs_ss_TEqs_001b},
         in wall-units \cite[(A3), p. 414]{Gerolymos_Vallet_2016b}, for the particular case of plane channel flow \eqrefsatob{Eq_FAepsijBsTPCF_s_AppendixFDPCF_ss_MFS_001}
                                                                                                                                  {Eq_FAepsijBsTPCF_s_AppendixFDPCF_ss_MFS_003}.}
\label{Tab_FAepsijBsTPCF_s_WAs_002}
\rule{\textwidth}{.25pt}
\end{center}
\end{table}
%
\begin{figure}
\begin{center}
\begin{picture}(450,550)
\put(0,-10){\includegraphics[angle=0,width=360pt,bb=59 51 515 763]{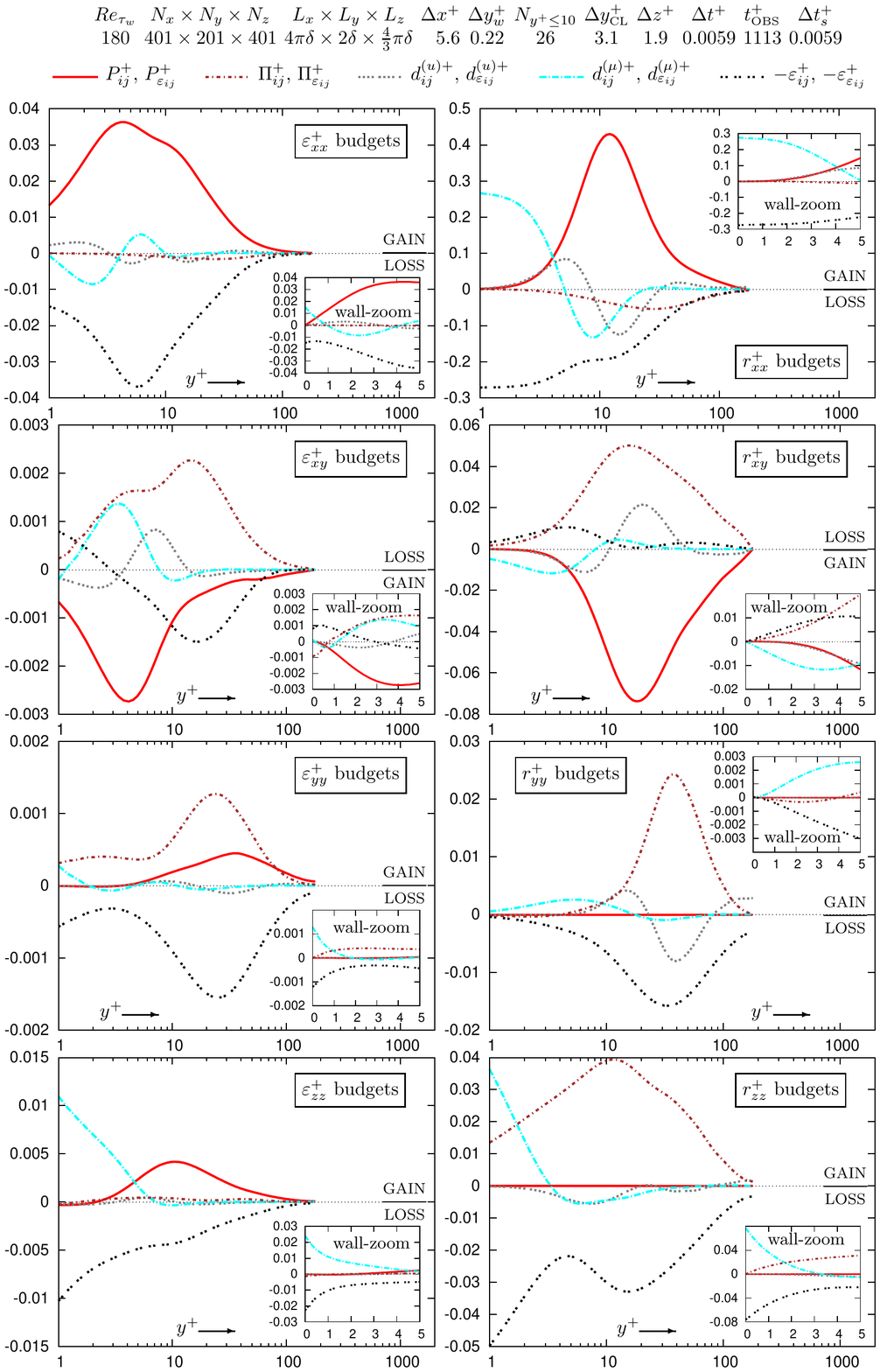}}
\end{picture}
\end{center}
\caption{Budgets, in wall-units \cite[(A3), p. 414]{Gerolymos_Vallet_2016b}, of the transport equations
for the dissipation tensor $\varepsilon_{ij}$ \eqref{Eq_FAepsijBsTPCF_s_TEqsWAs_ss_TEqs_001b} and for the Reynolds-stresses $r_{ij}$ \eqref{Eq_FAepsijBsTPCF_s_TEqsWAs_ss_TEqs_001a},
from the present \tsn{DNS} computations of turbulent plane channel flow ($Re_{\tau_w}\approxeq180$),
plotted against the inner-scaled wall-distance $y^+$ (logscale and linear wall-zoom).}
\label{Fig_FAepsijBsTPCF_s_TPCFBs_ss_epsijvsrijBs_001}
\end{figure}

%
%
%
%
%
%
%
%
%
\section{Turbulent plane channel flow budgets}\label{FAepsijBsTPCF_s_TPCFBs}
%
%
%
%
%
%
%
%
%

\tsn{DNS} data generated for plane channel flow \parref{FAepsijBsTPCF_s_TPCFBs_ss_DNSCs} illustrate how corresponding mechanisms in the transport equations of $r_{ij}$ \eqref{Eq_FAepsijBsTPCF_s_TEqsWAs_ss_TEqs_001a}
or $\varepsilon_{ij}$ \eqref{Eq_FAepsijBsTPCF_s_TEqsWAs_ss_TEqs_001b} contribute to the budgets of different components \parref{FAepsijBsTPCF_s_TPCFBs_ss_epsijvsrijBs}.
In direct analogy to $r_{ij}$-transport \cite{Mansour_Kim_Moin_1988a}, the fluctuating-pressure mechanisms in $\varepsilon_{ij}$-transport \eqref{Eq_FAepsijBsTPCF_s_TEqsWAs_ss_TEqs_001b}, $\Pi_{\varepsilon_{ij}}$,
can be analysed \parref{FAepsijBsTPCF_s_TPCFBs_ss_RPD} as the sum of a traceless redistributive term $\phi_{\varepsilon_{ij}}$ and a conservative pressure-diffusion part $d_{\varepsilon_{ij}}^{(u)}$.

%
%
%
%
%
\subsection{DNS computations}\label{FAepsijBsTPCF_s_TPCFBs_ss_DNSCs}
%
%
%
%
%

The \tsn{DNS} computations from which the present data were extracted are described in \citeasnoun{Gerolymos_Vallet_2016b}. They were obtained for low $Re_{\tau_w}\approxeq180$ plane channel
flow using a very-high-order \cite{Gerolymos_Senechal_Vallet_2009a} finite-volume solver \cite{Gerolymos_Senechal_Vallet_2010a} which has been thoroughly validated by comparison
with available \cite{Moser_Kim_Mansour_1999a,Hoyas_Jimenez_2008a,Vreman_Kuerten_2014a,Vreman_Kuerten_2014b,Vreman_Kuerten_2016a,Lee_Moser_2015a}
1-point and 2-point \tsn{DNS} data \cite{Gerolymos_Senechal_Vallet_2010a,Gerolymos_Senechal_Vallet_2013a,Gerolymos_Vallet_2014a,Gerolymos_Vallet_2016b}.

The terms in $\varepsilon_{ij}$-transport \eqref{Eq_FAepsijBsTPCF_s_TEqsWAs_ss_TEqs_001b} contain correlations of 1-order-higher derivatives of fluctuating quantities compared to the corresponding
terms in $r_{ij}$-transport \eqref{Eq_FAepsijBsTPCF_s_TEqsWAs_ss_TEqs_001a}. Therefore, terms in the $\varepsilon_{ij}$-transport equations \eqref{Eq_FAepsijBsTPCF_s_TEqsWAs_ss_TEqs_001b} are more sensitive
to computational truncation errors \cite{Gerolymos_2011a}, requiring finer grids to achieve the same accuracy as the corresponding terms in the $r_{ij}$-transport equations \eqref{Eq_FAepsijBsTPCF_s_TEqsWAs_ss_TEqs_001a}.
Furthermore, scaling analysis \cite[pp. 88--92]{Tennekes_Lumley_1972a} substantiates that terms in $\varepsilon_{ij}$-transport \eqref{Eq_FAepsijBsTPCF_s_TEqsWAs_ss_TEqs_001b} are generally
related with Taylor-microscale and/or Kolmogorov-scale structures, again suggesting that finer grids are required to obtain these terms than $\varepsilon_{ij}$ itself.
Accordingly, the computational grid resolution \figrefsab{Fig_FAepsijBsTPCF_s_TPCFBs_ss_epsijvsrijBs_001}
                                                         {Fig_FAepsijBsTPCF_s_TPCFBs_ss_RPD_001}
was high both streamwise ($\Delta x^+\approxeq5.6$) and spanwise ($\Delta z^+\approxeq1.9$)
to correctly predict the details of the elongated near-wall structures \cite[Figs. 12--15, pp. 802--805]{Gerolymos_Senechal_Vallet_2010a}.
Finally, several of the terms in $\varepsilon_{ij}$-transport \eqref{Eq_FAepsijBsTPCF_s_TEqsWAs_ss_TEqs_001b} present important variations in the viscous sublayer ($0<y^+\lessapprox3$; \figrefnp{Fig_FAepsijBsTPCF_s_TPCFBs_ss_epsijvsrijBs_001}),
requiring a fine wall-normal grid, not only at the wall ($\Delta y_w^+\approxeq0.22$ was found sufficient), but with weak cell-size stretching to ensure good resolution in the entire near-wall region
($N_{y^+\leq10}=26$ points in the region $0\leq y^+<10$) and actually throughout the entire channel up to the centerline ($\Delta y_\tsn{CL}^+\approxeq3.1$).
The streamwise resolution is similar to the finest grid used in \citeasnoun{Vreman_Kuerten_2016a} while the present spanwise resolution is roughly twice finer. On the other hand,
the present wall-normal resolution is roughly twice coarser compared to \citeasnoun{Vreman_Kuerten_2016a}. Although \citeasnoun{Vreman_Kuerten_2016a} did not study the dissipation tensor,
their data include the terms in the transport-equations for the variances of the fluctuating velocity-gradients \cite{Vreman_Kuerten_2014b},
which can be combined \cite{Gerolymos_Vallet_2016b} to obtain the transport equations for the diagonal terms $\{\varepsilon_{xx},\varepsilon_{yy},\varepsilon_{zz}\}$ (but not for the shear term $\varepsilon_{xy}$).
The 2 sets of data are in very good agreement \cite[Figs. 8, 9, pp. 410, 411]{Gerolymos_Vallet_2016b}.

Correlations in \eqref{Eq_FAepsijBsTPCF_s_TEqsWAs_ss_TEqs_001b} were computed using order-4 inhomogeneous-grid interpolating polynomials \cite{Gerolymos_2012b}
and sampled at every iteration ($\Delta t_s^+=\Delta t^+\approxeq0.0059$) for an observation interval $t_\tsn{OBS}^+\approxeq1113$. Because of the relatively short observation interval,
the pressure term $\Pi_{\varepsilon_{ij}}$ \eqref{Eq_FAepsijBsTPCF_s_TEqsWAs_ss_TEqs_001b} which contains the highly intermittent pressure-Hessian \cite[Fig. 12, p. 21]{Vreman_Kuerten_2014b},
was calculated from the identity $\Pi_{\varepsilon_{ij}}=d^{(p)}_{\varepsilon_{ij}}+\phi_{\varepsilon_{ij}}$ \eqref{Eq_FAepsijBsTPCF_s_TPCFBs_ss_RPD_001}.
The \tsn{RHS} terms in \eqref{Eq_FAepsijBsTPCF_s_TPCFBs_ss_RPD_001} only involve fluctuating pressure-gradients and converge much faster.

%
%
%
%
%
\subsection{$\varepsilon_{ij}$ {\em vs} $r_{ij}$ budgets}\label{FAepsijBsTPCF_s_TPCFBs_ss_epsijvsrijBs}
%
%
%
%
%

Comparison \figref{Fig_FAepsijBsTPCF_s_TPCFBs_ss_epsijvsrijBs_001} of the budgets of the Reynolds-stresses $r_{ij}$ \eqref{Eq_FAepsijBsTPCF_s_TEqsWAs_ss_TEqs_001a}
with those of the dissipation tensor $\varepsilon_{ij}$ \eqref{Eq_FAepsijBsTPCF_s_TEqsWAs_ss_TEqs_001b},
for plane channel flow \parref{FAepsijBsTPCF_s_AppendixFDPCF_ss_epsijBsPCF},
reveals fundamental differences, both in the relative importance of various mechanisms in the budgets of each component and in the componentality of corresponding
mechanisms.

Regarding the importance of different mechanisms in the budgets, it is noticeable that the pressure term $\Pi^+_{\varepsilon_{ij}}$
is negligibly small both for the streamwise $\varepsilon^+_{xx}$ and the spanwise $\varepsilon^+_{zz}$ components \figref{Fig_FAepsijBsTPCF_s_TPCFBs_ss_epsijvsrijBs_001}.
This difference is especially important in the budgets of the spanwise components, $r^+_{zz}$ and $\varepsilon^+_{zz}$. For the spanwise stress $r^+_{zz}$, in plane channel flow \eqrefsatob{Eq_FAepsijBsTPCF_s_AppendixFDPCF_ss_MFS_001}
                                                                                                                                                                                             {Eq_FAepsijBsTPCF_s_AppendixFDPCF_ss_MFS_003}
there is no production mechanism ($P^+_{zz}=0\;\forall\;y^+$) and gain comes mainly from the redistributive action of $\Pi^+_{ij}$ \figref{Fig_FAepsijBsTPCF_s_TPCFBs_ss_epsijvsrijBs_001}.
On the contrary, for the spanwise dissipation $\varepsilon^+_{zz}$, gain comes mainly from the production terms $P_{\varepsilon_{zz}}^{(2)+}+P_{\varepsilon_{zz}}^{(4)+}$ \eqref{Eq_FAepsijBsTPCF_s_AppendixFDPCF_ss_epsijBsPCF_002d},
the pressure term $\Pi^+_{\varepsilon_{zz}}$ being very weak \figref{Fig_FAepsijBsTPCF_s_TPCFBs_ss_epsijvsrijBs_001}.
\begin{figure}[h!]
\begin{center}
\begin{picture}(450,150)
\put( 0,-15){\includegraphics[angle=0,width=370pt,bb=60 460 515 659]{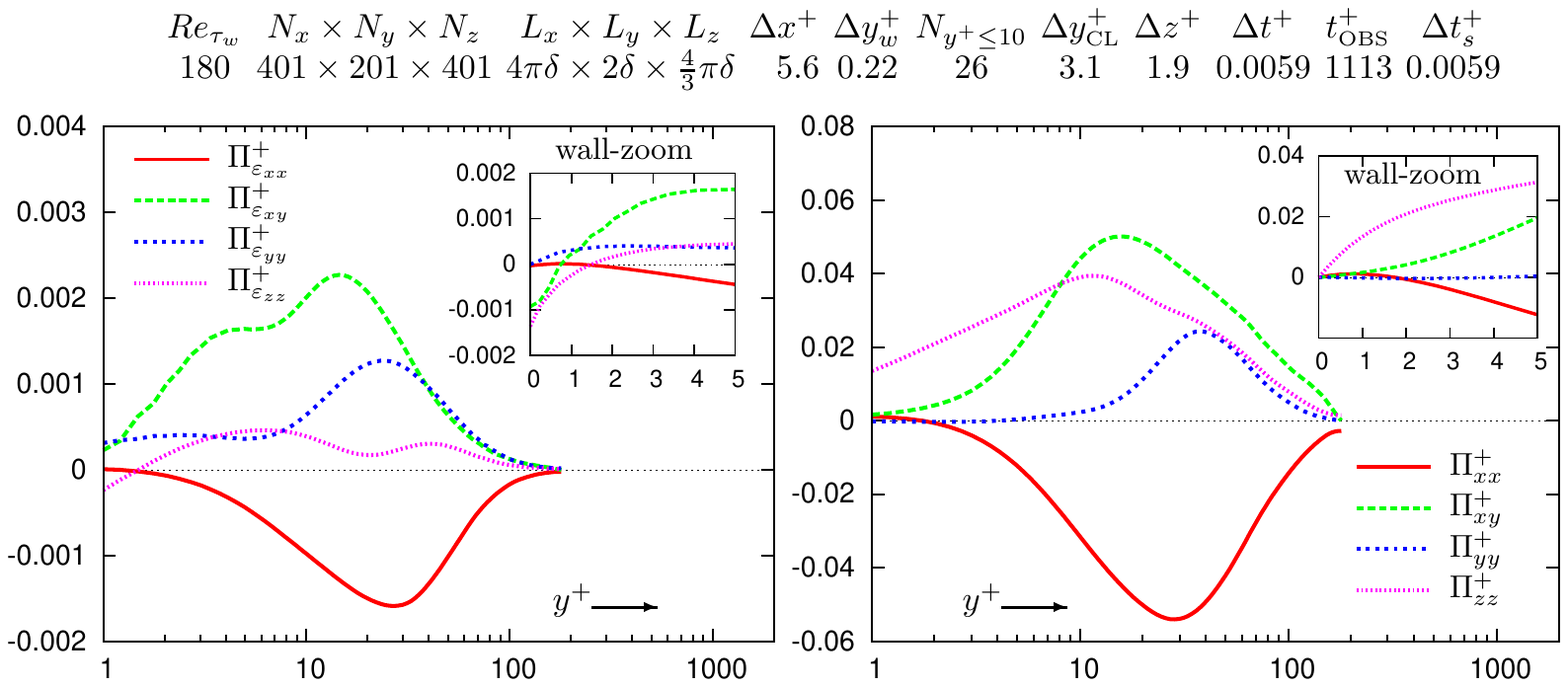}}
\end{picture}
\end{center}
\caption{Components, in wall-units \cite[(A3), p. 414]{Gerolymos_Vallet_2016b}, of the pressure terms $\Pi_{\varepsilon_{ij}}$ and $\Pi_{ij}$ in the transport equations
for the dissipation tensor $\varepsilon_{ij}$ \eqref{Eq_FAepsijBsTPCF_s_TEqsWAs_ss_TEqs_001b} and for the Reynolds-stresses $r_{ij}$ \eqref{Eq_FAepsijBsTPCF_s_TEqsWAs_ss_TEqs_001a},
from the present \tsn{DNS} computations of turbulent plane channel flow ($Re_{\tau_w}\approxeq180$),
plotted against the inner-scaled wall-distance $y^+$ (logscale and linear wall-zoom).}
\label{Fig_FAepsijBsTPCF_s_TPCFBs_ss_epsijvsrijBs_002}
%
\begin{center}
\begin{picture}(450,150)
\put( 0,-15){\includegraphics[angle=0,width=370pt,bb=59 460 515 659]{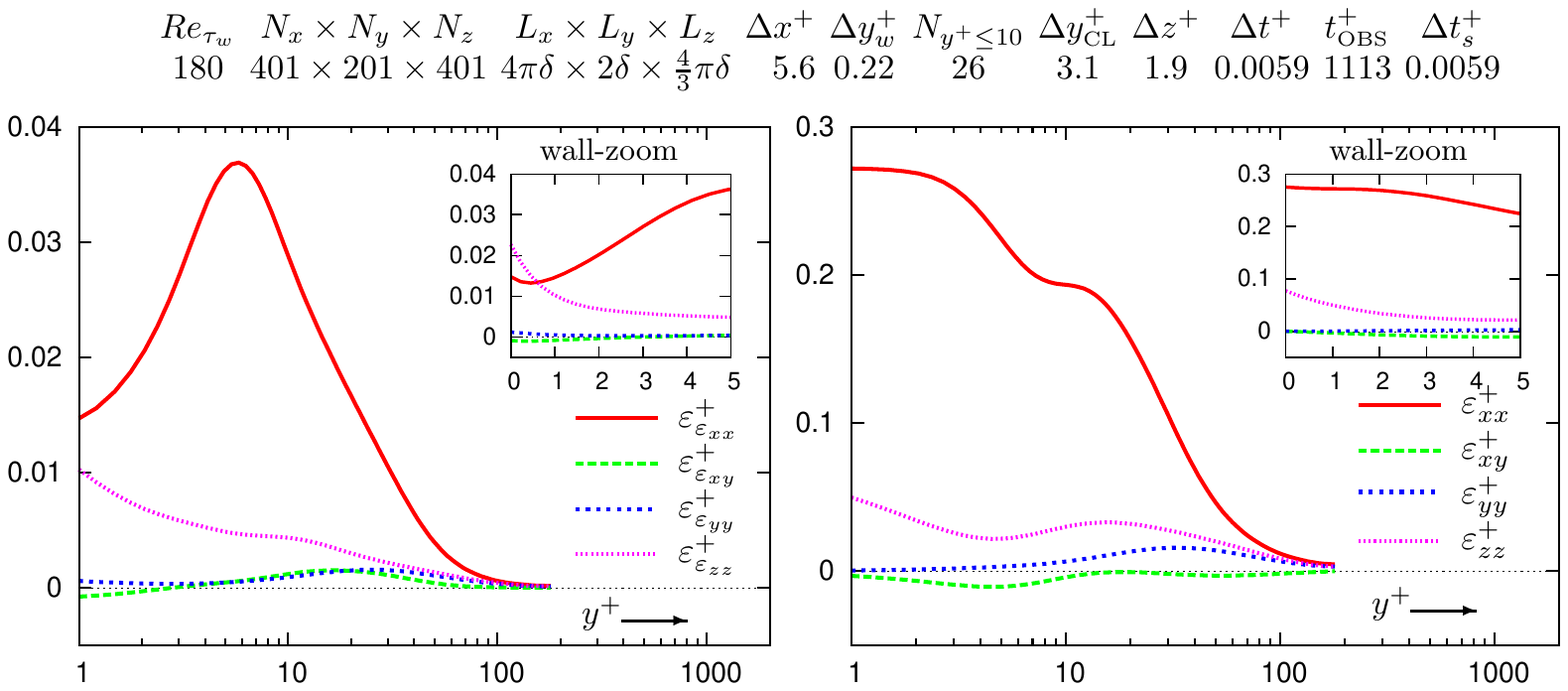}}
\end{picture}
\end{center}
\caption{Components, in wall-units \cite[(A3), p. 414]{Gerolymos_Vallet_2016b}, of the terms representing destruction by molecular viscosity $\varepsilon_{\varepsilon_{ij}}$ and $\varepsilon_{ij}$ in the transport equations
for the dissipation tensor $\varepsilon_{ij}$ \eqref{Eq_FAepsijBsTPCF_s_TEqsWAs_ss_TEqs_001b} and for the Reynolds-stresses $r_{ij}$ \eqref{Eq_FAepsijBsTPCF_s_TEqsWAs_ss_TEqs_001a},
from the present \tsn{DNS} computations of turbulent plane channel flow ($Re_{\tau_w}\approxeq180$),
plotted against the inner-scaled wall-distance $y^+$ (logscale and linear wall-zoom).}
\label{Fig_FAepsijBsTPCF_s_TPCFBs_ss_epsijvsrijBs_003}
\end{figure}
Comparison of the componentality of $\Pi^+_{\varepsilon_{ij}}$ with that of $\Pi^+_{ij}$ \figref{Fig_FAepsijBsTPCF_s_TPCFBs_ss_epsijvsrijBs_002} reveals that, although
all the components of each tensor are of the same order-of-magnitude, $\Pi^+_{\varepsilon_{zz}}$ is consistently weaker than the other components of $\Pi^+_{\varepsilon_{ij}}$
contrary to $\Pi^+_{zz}$ which is the largest component of $\Pi^+_{ij}$ near the wall ($y^+\lessapprox10$; \figrefnp{Fig_FAepsijBsTPCF_s_TPCFBs_ss_epsijvsrijBs_002}).
Another important difference is observed in the limiting behaviour of $\Pi^+_{\varepsilon_{xy}}$ and $\Pi^+_{\varepsilon_{zz}}$ both of which are $\neq0$ at the wall \tabref{Tab_FAepsijBsTPCF_s_WAs_001}
whereas $[\Pi_{ij}]^+_w=0$ because of the no-slip condition \eqref{Eq_FAepsijBsTPCF_s_AppendixFDPCF_ss_MFS_001a}.

The $y^+$-distribution \figref{Fig_FAepsijBsTPCF_s_TPCFBs_ss_epsijvsrijBs_003} of the destruction-of-dissipation tensor $\varepsilon^+_{\varepsilon_{ij}}$ \eqref{Eq_FAepsijBsTPCF_s_TEqsWAs_ss_TEqs_001b}
differs substantially from that of the dissipation tensor $\varepsilon^+_{ij}$ \eqref{Eq_FAepsijBsTPCF_s_TEqsWAs_ss_TEqs_001a}.
Away from the wall, the streamwise components $\varepsilon^+_{xx}$ and $\varepsilon^+_{\varepsilon_{xx}}$ are in both cases much larger than the other components.
Near the wall $\varepsilon^+_{xx}$ forms a small plateau ($y^+\in[8,12]$; \figrefnp{Fig_FAepsijBsTPCF_s_TPCFBs_ss_epsijvsrijBs_003}) and then increases as $y^+\to 0$,
reaching its global maximum at the wall, remaining by far the largest component of $\varepsilon^+_{ij}\;\forall\;y^+$ \figref{Fig_FAepsijBsTPCF_s_TPCFBs_ss_epsijvsrijBs_003}.
On the contrary, $\varepsilon^+_{\varepsilon_{xx}}$ reaches its global maximum at $y^+\approxeq 7$ and then decreases as $y^+\to 0$. At the same time $\varepsilon^+_{\varepsilon_{zz}}$ sharply increases
near the wall, the 2 components crossing each other at $y^+\approxeq 0.7$ \figref{Fig_FAepsijBsTPCF_s_TPCFBs_ss_epsijvsrijBs_003} to reach $[\varepsilon_{\varepsilon_{zz}}^+]^+_w>[\varepsilon_{\varepsilon_{xx}}^+]^+_w$.
The wall-asymptotic expansion of $\varepsilon^+_{\varepsilon_{ij}}$, as $y^+\to 0$, shows \tabref{Tab_FAepsijBsTPCF_s_WAs_002}
that all of the $\varepsilon^+_{\varepsilon_{ij}}$-components are $\neq0$ at the wall
in contrast to $\varepsilon_{ij}^+$, for which $[\varepsilon_{yy}]^+_w=[\varepsilon_{xy}]^+_w=0$ \cite[(16,21), pp. 21--22]{Mansour_Kim_Moin_1988a}.
Another difference in the componentality of the 2 tensors \figref{Fig_FAepsijBsTPCF_s_TPCFBs_ss_epsijvsrijBs_003} is that while $\varepsilon_{xy}^+<0 \;\forall\;y^+\in\;]0,\delta^+[$,
$\varepsilon_{\varepsilon_{xy}}^+\leq 0\;\forall\;y^+\lessapprox 3$ changes sign further away from 
the wall ($\varepsilon_{\varepsilon_{xy}}^+\geq 0\;\forall\;y^+\gtrapprox 3$; \figrefnp{Fig_FAepsijBsTPCF_s_TPCFBs_ss_epsijvsrijBs_003}).
Therefore, while $-\varepsilon_{xy}^+>0 \;\forall\;y^+\in\;]0,\delta^+[$ is a loss mechanism in the budgets of $r_{xy}^+<0 \;\forall\;y^+\in\;]0,\delta^+[$ \figref{Fig_FAepsijBsTPCF_s_TPCFBs_ss_epsijvsrijBs_001},
this is not the case for $-\varepsilon_{\varepsilon_{xy}}^+$ which is, in the major part of the channel ($y^+\gtrapprox3$; \figrefnp{Fig_FAepsijBsTPCF_s_TPCFBs_ss_epsijvsrijBs_001}),
a gain mechanism in the $\varepsilon_{xy}$-budgets.
The componentality differences between $r_{ij}$, its dissipation $\varepsilon_{ij}$ and the destruction-of-dissipation $\varepsilon_{\varepsilon_{ij}}$ are further studied in \parrefnp{FAepsijBsTPCF_s_epsepsij}.

The most stricking componentality difference concerns the production mechanisms, $P^+_{ij}$ \eqref{Eq_FAepsijBsTPCF_s_TEqsWAs_ss_TEqs_001a} and $P^+_{\varepsilon_{ij}}$ \eqref{Eq_FAepsijBsTPCF_s_TEqsWAs_ss_TEqs_001b}.
In plane channel flow, all of the components of $P^+_{\varepsilon_{ij}}$ are generally $\neq0$ and contribute as gain to the corresponding $\varepsilon^+_{ij}$ component \figref{Fig_FAepsijBsTPCF_s_TPCFBs_ss_epsijvsrijBs_001},
contrary to $P_{ij}$ \citeaffixed{Mansour_Kim_Moin_1988a}{in plane channel flow $P_{yy}=P_{zz}=0\;\forall\;y^+$}.
The production mechanisms \eqref{Eq_FAepsijBsTPCF_s_TEqsWAs_ss_TEqs_001b} $P_{\varepsilon_{ij}}^{(1)+}$ (by the direct action of the components of $\varepsilon^+_{ij}$ on the mean velocity-gradient)
and $P_{\varepsilon_{ij}}^{(3)+}$ (related to the mean velocity-Hessian) have a similar componentality
($P_{\varepsilon_{yy}}^{(1)+}=P_{\varepsilon_{zz}}^{(1)+}=P_{\varepsilon_{yy}}^{(3)+}=P_{\varepsilon_{zz}}^{(3)+}=0\;\forall\;y^+$) in plane channel flow \eqrefsatob{Eq_FAepsijBsTPCF_s_AppendixFDPCF_ss_MFS_001}
                                                                                                                                                                     {Eq_FAepsijBsTPCF_s_AppendixFDPCF_ss_MFS_003},
but this is not the case for the second production by mean velocity-gradient mechanism $P_{\varepsilon_{ij}}^{(2)+}$ nor for the production by the triple correlations of fluctuating velocity-gradients $P_{\varepsilon_{ij}}^{(4)+}$,
both of which are generally $\neq0$ for all of the components \cite[Fig. 6, p. 407]{Gerolymos_Vallet_2016b}.

At the wall ($y^+=0$), production $P^+_{\varepsilon_{ij}}$ and turbulent diffusion by the fluctuating velocities $d^{(u)+}_{\varepsilon_{ij}}$ are $0$
\begin{subequations}
                                                                                                                                    \label{Eq_DTWT_s_epsijBs_ss_epsijTWAs_001}
\begin{alignat}{6}
\tabrefsab{Tab_FAepsijBsTPCF_s_WAs_001}
          {Tab_FAepsijBsTPCF_s_WAs_002}\implies[P_{\varepsilon_{ij}}]^+_w=[d^{(u)}_{\varepsilon_{ij}}]^+_w=0
                                                                                                                                    \label{Eq_DTWT_s_epsijBs_ss_epsijTWAs_001a}
\end{alignat}
so that the wall-budgets of the $\varepsilon_{ij}$-transport equations \eqrefsab{Eq_FAepsijBsTPCF_s_AppendixFDPCF_ss_epsijBsPCF_001}
                                                                                {Eq_FAepsijBsTPCF_s_AppendixFDPCF_ss_epsijBsPCF_002} reduce to
\begin{alignat}{6}
\tabrefsab{Tab_FAepsijBsTPCF_s_WAs_001}
          {Tab_FAepsijBsTPCF_s_WAs_002}\implies[d^{(\mu)}_{\varepsilon_{ij}}]^+_w+[\Pi_{\varepsilon_{ij}}]^+_w=[\varepsilon_{\varepsilon_{ij}}]^+_w
                                                                                                                                    \label{Eq_DTWT_s_epsijBs_ss_epsijTWAs_001b}
\end{alignat}
\end{subequations}
In the particular case of the wall-normal diagonal component $[\Pi_{\varepsilon_{yy}}]^+_w=0$ \tabref{Tab_FAepsijBsTPCF_s_WAs_001},
implying $[d^{(\mu)}_{\varepsilon_{yy}}]^+_w=[\varepsilon_{\varepsilon_{yy}}]^+_w=16\overline{{B_v'^+}^2}$ \tabrefsab{Tab_FAepsijBsTPCF_s_WAs_001}
                                                                                                                     {Tab_FAepsijBsTPCF_s_WAs_002}.
Notice also that, by \eqref{Eq_FAepsijBsTPCF_s_AppendixABVSy+0_ss_FCEq_004},
the halftrace $\tfrac{1}{2}[\Pi_{\varepsilon_{\ell\ell}}]^+_w\stackrel{\tabref{Tab_FAepsijBsTPCF_s_WAs_001}}{=}-8\overline{{B_v'^+}^2}$
in agreement with \citeasnoun[(24), p. 24]{Mansour_Kim_Moin_1988a}.
\begin{figure}[h!]
\begin{center}
\begin{picture}(450,310)
\put(0,-15){\includegraphics[angle=0,width=370pt,bb=60 380 515 763]{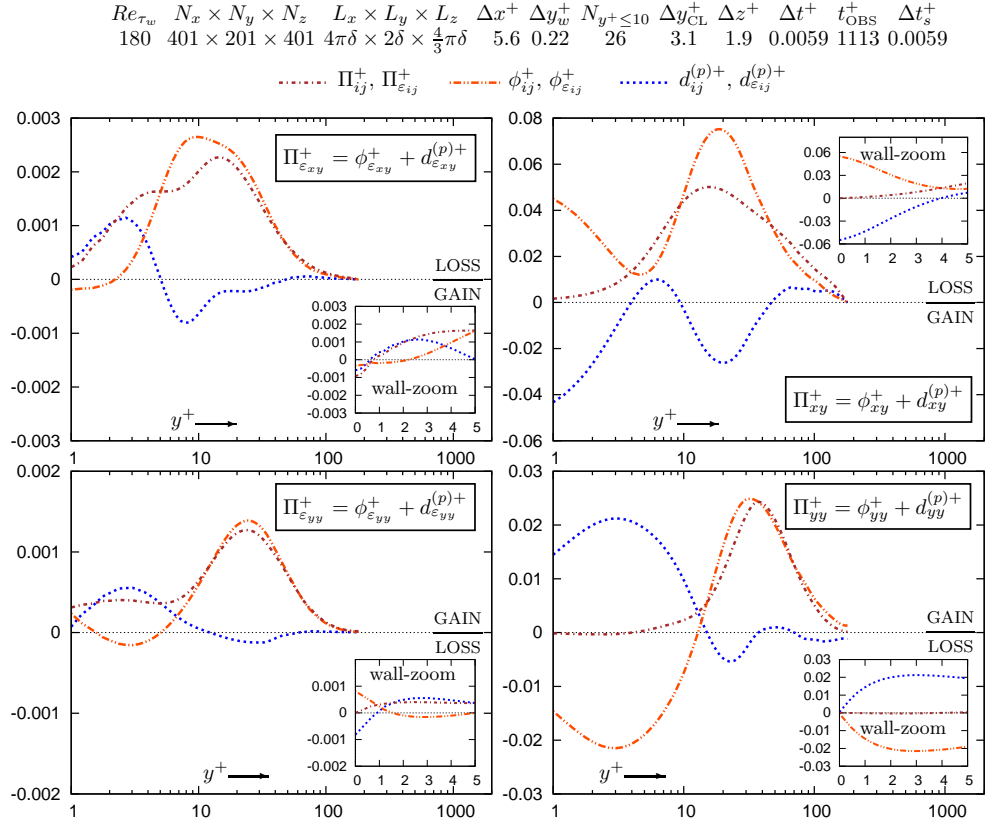}}
\end{picture}
\end{center}
\caption{Splitting \eqrefsab{Eq_FAepsijBsTPCF_s_TPCFBs_ss_RPD_001}
                            {Eq_FAepsijBsTPCF_s_TPCFBs_ss_RPD_002},
of wall-normal and shear components (in plane channel flow $d_{\varepsilon_{xx}}^{(p)+}=d_{\varepsilon_{zz}}^{(p)+}=d_{xx}^{(p)+}=d_{zz}^{(p)+}=0\;\forall y^+$)
of the pressure terms $\Pi_{\varepsilon_{ij}}$ and $\Pi_{ij}$ in the transport equations
for the dissipation tensor $\varepsilon_{ij}$ \eqref{Eq_FAepsijBsTPCF_s_TEqsWAs_ss_TEqs_001b} and for the Reynolds-stresses $r_{ij}$ \eqref{Eq_FAepsijBsTPCF_s_TEqsWAs_ss_TEqs_001a},
into redistribution ($\Pi_{\varepsilon_{ij}}$ and $\Pi_{ij}$) and pressure diffusion ($d_{\varepsilon_{ij}}^{(p)}$ and $d_{ij}^{(p)}$),
in wall-units \cite[(A3), p. 414]{Gerolymos_Vallet_2016b},
from the present \tsn{DNS} computations of turbulent plane channel flow ($Re_{\tau_w}\approxeq180$),
plotted against the inner-scaled wall-distance $y^+$ (logscale and linear wall-zoom).}
\label{Fig_FAepsijBsTPCF_s_TPCFBs_ss_RPD_001}
\end{figure}

%
%
%
%
%
\subsection{Redistribution and pressure-diffusion}\label{FAepsijBsTPCF_s_TPCFBs_ss_RPD}
%
%
%
%
%

In exact analogy with $r_{ij}$-transport \eqref{Eq_FAepsijBsTPCF_s_TEqsWAs_ss_TEqs_001a}, where by application of the product-rule of differentiation \cite[\S2.12, pp. 44-46]{Riley_Hobson_Bence_2006a},
the velocity/pressure-gradient correlation $\Pi_{ij}$ \eqref{Eq_FAepsijBsTPCF_s_TEqsWAs_ss_TEqs_001a} can be split into pressure diffusion $d^{(p)}_{ij}$ and a redistributive term $\phi_{ij}$
\begin{subequations}
                                                                                                                                    \label{Eq_FAepsijBsTPCF_s_TPCFBs_ss_RPD_001}
\begin{alignat}{6}
\Pi_{ij}\stackrel{\eqref{Eq_FAepsijBsTPCF_s_TEqsWAs_ss_TEqs_001a}}{=}\underbrace{\dfrac{\partial}{\partial  x_\ell}\left(-\delta_{i\ell}\overline{u'_j p'}
                                                                                                             -\delta_{j\ell}\overline{u'_i p'}\right)}_{\displaystyle d^{(p)}_{ij}}
                                                        +\underbrace{\overline{p'\left(\dfrac{\partial u'_i}
                                                                                             {\partial  x_j}+\dfrac{\partial u'_j}
                                                                                                                   {\partial  x_i}\right)}}_{\displaystyle \phi_{ij}}
                                                                                                                                    \label{Eq_FAepsijBsTPCF_s_TPCFBs_ss_RPD_001a}
\end{alignat}
with
\begin{alignat}{6}
\phi_{mm}\stackrel{\eqrefsab{Eq_FAepsijBsTPCF_s_TPCFBs_ss_RPD_001a}
                            {Eq_FAepsijBsTPCF_s_AppendixABVSy+0_ss_FCEq_002}}{=}0\stackrel{\eqref{Eq_FAepsijBsTPCF_s_TPCFBs_ss_RPD_001a}}{\implies}
\Pi_{mm}=d^{(p)}_{mm}\stackrel{\eqref{Eq_FAepsijBsTPCF_s_TPCFBs_ss_RPD_001a}}{=}\dfrac{\partial}{\partial  x_\ell}\left(-2\overline{u'_\ell p'}\right)
                                                                                                                                    \label{Eq_FAepsijBsTPCF_s_TPCFBs_ss_RPD_001b}
\end{alignat}
\end{subequations}
the pressure term $\Pi_{\varepsilon_{ij}}$ in \eqref{Eq_FAepsijBsTPCF_s_TEqsWAs_ss_TEqs_001b} can be split into pressure diffusion $d^{(p)}_{\varepsilon_{ij}}$ and a redistributive term $\phi_{\varepsilon_{ij}}$, \viz
\begin{subequations}
                                                                                                                                    \label{Eq_FAepsijBsTPCF_s_TPCFBs_ss_RPD_002}
\begin{alignat}{6}
\Pi_{\varepsilon_{ij}}\stackrel{\eqref{Eq_FAepsijBsTPCF_s_TEqsWAs_ss_TEqs_001b}}{=}
                      \underbrace{\dfrac{\partial}{\partial  x_\ell}\left(-2\nu\delta_{i\ell}\overline{\dfrac{\partial u'_j}
                                                                                                             {\partial  x_k}\dfrac{\partial   p'}
                                                                                                                                  {\partial  x_k}}
                                                                          -2\nu\delta_{j\ell}\overline{\dfrac{\partial u'_i}
                                                                                                             {\partial  x_k}\dfrac{\partial   p'}
                                                                                                                                  {\partial  x_k}}\right)}_{\displaystyle d^{(p)}_{\varepsilon_{ij}}}
                     +\underbrace{2\nu\overline{\dfrac{\partial   p'}
                                                      {\partial  x_k}\left[\dfrac{\partial}{\partial  x_k}\left(\dfrac{\partial u'_i}
                                                                                                                      {\partial  x_j}+\dfrac{\partial u'_j}
                                                                                                                                            {\partial  x_i}\right)\right]}}_{\displaystyle \phi_{\varepsilon_{ij}}}
                                                                                                                                    \label{Eq_FAepsijBsTPCF_s_TPCFBs_ss_RPD_002a}
\end{alignat}
with
\begin{alignat}{6}
\phi_{\varepsilon_{mm}}\stackrel{\eqrefsab{Eq_FAepsijBsTPCF_s_TPCFBs_ss_RPD_002a}
                                          {Eq_FAepsijBsTPCF_s_AppendixABVSy+0_ss_FCEq_002}}{=}0\stackrel{\eqref{Eq_FAepsijBsTPCF_s_TPCFBs_ss_RPD_002a}}{\implies}
\Pi_{\varepsilon_{mm}}=d^{(p)}_{\varepsilon_{mm}}\stackrel{\eqref{Eq_FAepsijBsTPCF_s_TPCFBs_ss_RPD_002a}}{=}\dfrac{\partial}{\partial  x_\ell}\left(-4\nu\overline{\dfrac{\partial u'_\ell}
                                                                                                                                                                              {\partial  x_k}\dfrac{\partial   p'}
                                                                                                                                                                                                   {\partial  x_k}}\right)
                                                                                                                                    \label{Eq_FAepsijBsTPCF_s_TPCFBs_ss_RPD_002b}
\end{alignat}
\end{subequations}
Because of the incompressible fluctuating continuity \eqref{Eq_FAepsijBsTPCF_s_AppendixABVSy+0_ss_FCEq_002},
$\phi_{\varepsilon_{ij}}$ \eqref{Eq_FAepsijBsTPCF_s_TPCFBs_ss_RPD_002a} is traceless \eqref{Eq_FAepsijBsTPCF_s_TPCFBs_ss_RPD_002b},
exactly like $\phi_{ij}$ \eqref{Eq_FAepsijBsTPCF_s_TPCFBs_ss_RPD_001}. Therefore it does not appear in the transport equation for the dissipation-rate $\varepsilon$ of the turbulence kinetic energy
\cite[(23), p. 23]{Mansour_Kim_Moin_1988a} and has a redistribution role among components of $\varepsilon_{ij}$. In second-moment closures, $\phi_{ij}$ \eqref{Eq_FAepsijBsTPCF_s_TPCFBs_ss_RPD_001}
occupies a central place \cite{Launder_Reece_Rodi_1975a,Speziale_Sarkar_Gatski_1991a,Gerolymos_Lo_Vallet_2012a,Jakirlic_Hanjalic_2013a}
in modelling work, because pressure diffusion $d^{(p)}_{ij}$ is absent in homogeneous flows. It is therefore interesting to investigate \figref{Fig_FAepsijBsTPCF_s_TPCFBs_ss_RPD_001}
the splitting \eqref{Eq_FAepsijBsTPCF_s_TPCFBs_ss_RPD_002} of $\Pi_{\varepsilon_{ij}}$ in comparison with the splitting of $\Pi_{ij}$ \eqref{Eq_FAepsijBsTPCF_s_TPCFBs_ss_RPD_001}.
Since only $y$-gradients of second-moments of fluctuating quantities are $\neq0$ in plane channel flow \eqref{Eq_FAepsijBsTPCF_s_AppendixFDPCF_ss_MFS_003}
the splittings \eqrefsab{Eq_FAepsijBsTPCF_s_TPCFBs_ss_RPD_001}
                        {Eq_FAepsijBsTPCF_s_TPCFBs_ss_RPD_002}
are only relevant for the wall-normal and the shear components (in plane channel flow $d_{\varepsilon_{xx}}^{(p)+}=d_{\varepsilon_{zz}}^{(p)+}=d_{xx}^{(p)+}=d_{zz}^{(p)+}=0\;\forall\;y^+$).

As already observed in the analysis of $r_{ij}$-transport \cite{Mansour_Kim_Moin_1988a}, pressure diffusion is generally weak away from the wall,
so that \figref{Fig_FAepsijBsTPCF_s_TPCFBs_ss_RPD_001}
both $\Pi_{\varepsilon_{yy}}\approxeq\phi_{\varepsilon_{yy}}\;\forall\;y^+\gtrapprox10$ and $\Pi_{yy}\approxeq\phi_{yy}\;\forall\;y^+\gtrapprox10$.
These approximate equalities also apply for the shear components, but for higher $y^+\gtrapprox30$ \figref{Fig_FAepsijBsTPCF_s_TPCFBs_ss_RPD_001}.
This implies that modelling $\phi_{\varepsilon_{ij}}$ in lieu of $\Pi_{\varepsilon_{ij}}$ in the log-region of the velocity profile \cite{Coles_1956a} could be a reasonable working choice,
exactly like in $r_{ij}$-transport models \cite{Launder_Reece_Rodi_1975a}. On the other hand, nearer to the wall ($1\lessapprox y^+\lessapprox10$; \figrefnp{Fig_FAepsijBsTPCF_s_TPCFBs_ss_RPD_001})
the splittings of $\Pi_{\varepsilon_{ij}}$ \eqref{Eq_FAepsijBsTPCF_s_TPCFBs_ss_RPD_002} and $\Pi_{ij}$ \eqref{Eq_FAepsijBsTPCF_s_TPCFBs_ss_RPD_001} are quite different.
Regarding $\Pi_{ij}$, both $\Pi_{yy}$ and $\Pi_{xy}$ are very small for $y^+\lessapprox5$, so that
$\phi_{yy}\approxeq-d^{(p)}_{yy}\;\forall y^+\lessapprox8$ and $\phi_{xy}\approxeq-d^{(p)}_{xy}\;\forall y^+\lessapprox4$ \figref{Fig_FAepsijBsTPCF_s_TPCFBs_ss_RPD_001},
but this dows not apply to $\Pi_{\varepsilon_{ij}}$.
Notice also that while $[\Pi_{ij}]_w\stackrel{\eqrefsab{Eq_FAepsijBsTPCF_s_TEqsWAs_ss_TEqs_001a}{Eq_FAepsijBsTPCF_s_AppendixABVSy+0_ss_FCEq_002}}{=}0$ because of the no-slip condition at the wall
this is not the case for $\Pi_{\varepsilon_{ij}}$ (only the wall-normal component $[\Pi_{\varepsilon_{yy}}]_w=0$ at the wall; \tabrefnp{Tab_FAepsijBsTPCF_s_WAs_001}).
These differences in near-wall behaviour between $\Pi_{\varepsilon_{ij}}$ and $\Pi_{ij}$ should be kept in mind in modelling efforts of the pressure terms in differential $\varepsilon_{ij}$-transport closures.
\begin{table}
\begin{center}
\rule{\textwidth}{.25pt}
\input{Tab_FluidDynRes_01_WAs_epseps}
\caption{Asymptotic (as $y^+\to0$) expansions of the components of $\tsr{\varepsilon_\varepsilon}$ \eqref{Eq_FAepsijBsTPCF_s_TEqsWAs_ss_TEqs_001b}, in wall-units \cite[(A3), p. 414]{Gerolymos_Vallet_2016b},
         for general inhomogeneous incompressible turbulent flow near a plane no-slip $xz$-wall
         (terms within square brackets {\color{blue}{$[\cdots]$}} are 3-D terms which are identically $=0$ for 2-D in-the-mean flow
         whereas the term within curly brackets {\color{blue}{$\{\cdots\}$}} in $\varepsilon_{xy}^+$,
         $\overline{B_u'^+C_v'^+}=0$ \eqref{Eq_FAepsijBsTPCF_s_AppendixABVSy+0_ss_PCF_sss_WPFM_003c} in the particular case of plane channel flow).}
\label{Tab_FAepsijBsTPCF_s_epsepsij_001}
\rule{\textwidth}{.25pt}
\end{center}
\end{table}
%
\begin{figure}
\begin{center}
\begin{picture}(450,500)
\put(0,-15){\includegraphics[angle=0,width=370pt,bb=81 135 509 708]{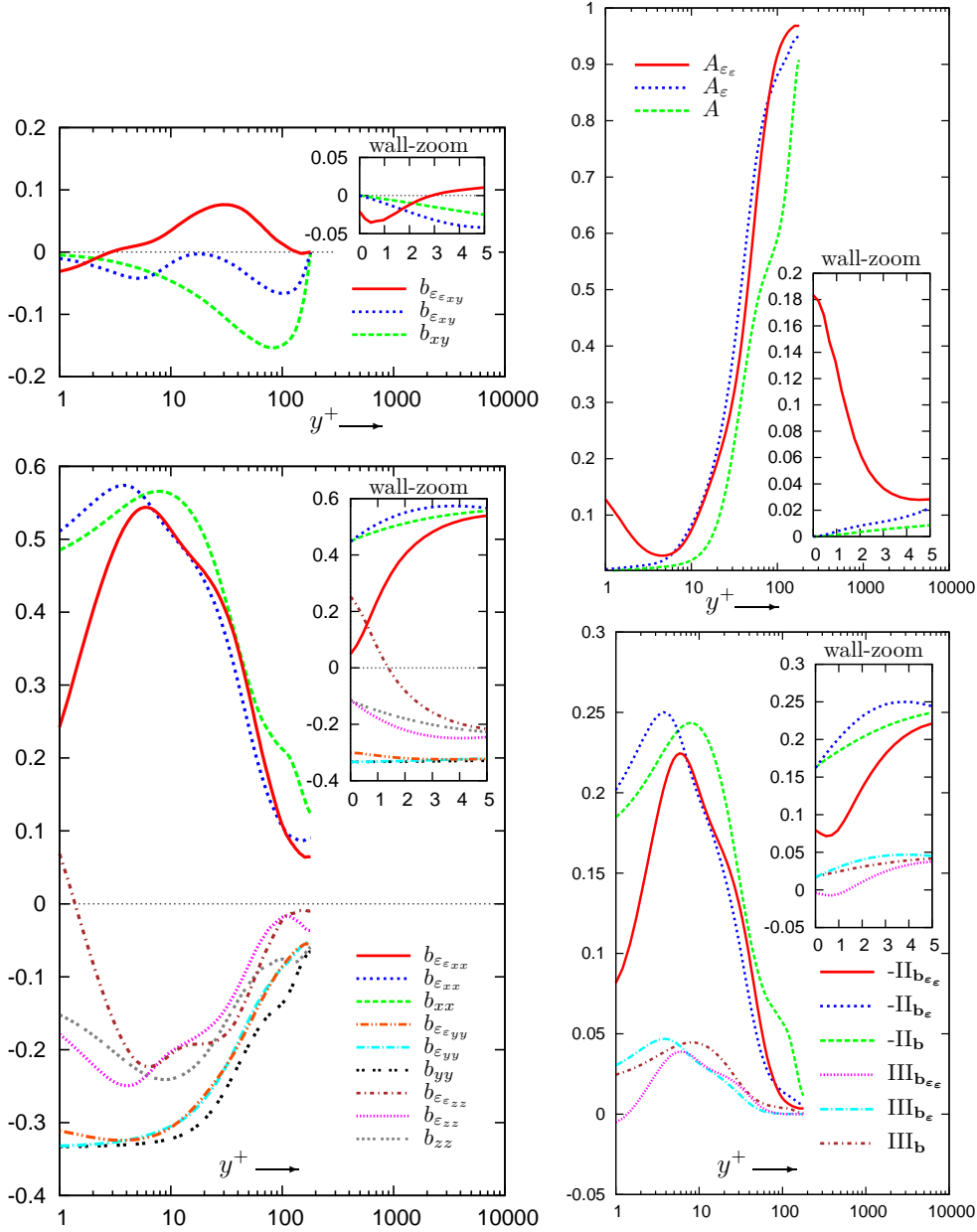}}
\end{picture}
\end{center}
\caption{Components and invariants of the anisotropy tensors  of the Reynolds-stresses $b_{ij}$ \eqref{Eq_FAepsijBsTPCF_s_epsepsij_002a},
of the dissipation tensor $b_{\varepsilon_{ij}}$ \eqref{Eq_FAepsijBsTPCF_s_epsepsij_002b} and of the destruction-of-dissipation tensor $b_{\varepsilon_{\varepsilon_{ij}}}$ \eqref{Eq_FAepsijBsTPCF_s_epsepsij_002c},
from the present \tsn{DNS} computations of turbulent plane channel flow ($Re_{\tau_w}\approxeq180$),
plotted against the inner-scaled wall-distance $y^+$ (logscale and linear wall-zoom).}
\label{Fig_FAepsijBsTPCF_s_epsepsij_001}
\end{figure}
%
\begin{sidewaystable}
\vspace{6.0in}
\begin{center}
\rule{\textwidth}{.25pt}
\input{Tab_FluidDynRes_01_WAs_bepseps}
\caption{Asymptotic (as $y^+\to0$) expansions of the components of $\tsr{b_{\varepsilon_\varepsilon}}$ \eqref{Eq_Eq_FAepsijBsTPCF_s_epsepsij_001c}, in wall-units \cite[(A3), p. 414]{Gerolymos_Vallet_2016b},
         for general inhomogeneous incompressible turbulent flow near a plane no-slip $xz$-wall
         (terms within square brackets {\color{blue}{$[\cdots]$}} are 3-D terms which are identically $=0$ for 2-D in-the-mean flow
         whereas the term within curly brackets {\color{blue}{$\{\cdots\}$}} in $b_{\varepsilon_{\varepsilon_{xy}}}$,
         $\overline{B_u'^+C_v'^+}=0$ \eqref{Eq_FAepsijBsTPCF_s_AppendixABVSy+0_ss_PCF_sss_WPFM_003c} in the particular case of plane channel flow).}
\label{Tab_FAepsijBsTPCF_s_epsepsij_002}
\rule{\textwidth}{.25pt}
\end{center}
\end{sidewaystable}

%
%
%
%
%
%
%
%
%
\section{Destruction-of-dissipation tensor $\varepsilon_{\varepsilon_{ij}}$}\label{FAepsijBsTPCF_s_epsepsij}
%
%
%
%
%
%
%
%
%

The diagonal components and traces of the 3 tensors
\begin{subequations}
                                                                                                                                    \label{Eq_FAepsijBsTPCF_s_epsepsij_001}
\begin{alignat}{6}
r_{ij}&:=&\overline{u_i'u_j'}\implies& r_{xx},r_{yy},r_{zz}&\geq0\leq&
\TKE:=\tfrac{1}{2}\overline{u_m'u_m'}
                                                                                                                                    \label{Eq_FAepsijBsTPCF_s_epsepsij_001b}\\
\varepsilon_{ij}&:=&2\nu\overline{\dfrac{\partial u_i'  }
                                      {\partial x_\ell}
                                \dfrac{\partial u_j'  }
                                      {\partial x_\ell}}\implies& \varepsilon_{xx},\varepsilon_{yy},\varepsilon_{zz}&\geq0\leq&
\varepsilon:=\tfrac{1}{2}\varepsilon_{mm}
                                                                                                                                    \label{Eq_FAepsijBsTPCF_s_epsepsij_001b}\\
\varepsilon_{\varepsilon_{ij}}&:=4&\nu^2\overline{\dfrac{\partial^2 u_i'            }
                                                      {\partial x_k\partial x_\ell}
                                                \dfrac{\partial^2 u_j'             }
                                                      {\partial x_k\partial x_\ell}}\implies& \varepsilon_{\varepsilon_{xx}},\varepsilon_{\varepsilon_{yy}},\varepsilon_{\varepsilon_{zz}}&\geq0\leq&
\varepsilon_{\varepsilon}:=\tfrac{1}{2}\varepsilon_{\varepsilon_{mm}}
                                                                                                                                    \label{Eq_Eq_FAepsijBsTPCF_s_epsepsij_001c}
\end{alignat}
\end{subequations}
are positive in every frame-of-reference. Therefore these tensors are positive-definite \cite{Gerolymos_Vallet_2016b}
implying that the invariants \cite{Rivlin_1955a} of the corresponding traceless anisotropy tensors \cite{Gerolymos_Lo_Vallet_2012a}
\begin{subequations}
                                                                                                                                    \label{Eq_FAepsijBsTPCF_s_epsepsij_002}
\begin{alignat}{6}
b_{ij}:=&\frac{\overline{u_i'u_j'}}{2\TKE}-\tfrac{1}{3}\delta_{ij}&&\quad;\quad
\II{b}=&&-\tfrac{1}{2}b_{mk}b_{km}&&\quad;\quad
\III{b}=&&\tfrac{1}{3}b_{mk}b_{k\ell}b_{\ell m}
                                                                                                                                    \label{Eq_FAepsijBsTPCF_s_epsepsij_002a}\\
b_{\varepsilon_{ij}}:=&\frac{\varepsilon_{ij}}{2\varepsilon}-\tfrac{1}{3}\delta_{ij}&&\quad;\quad
\II{b_\varepsilon}=&&-\tfrac{1}{2}b_{\varepsilon_{mk}}b_{\varepsilon_{km}}&&\quad;\quad
\III{b_\varepsilon}=&&\tfrac{1}{3}b_{\varepsilon_{mk}}b_{\varepsilon_{k\ell}}b_{\varepsilon_{\ell m}}
                                                                                                                                    \label{Eq_FAepsijBsTPCF_s_epsepsij_002b}\\
b_{\varepsilon_{\varepsilon_{ij}}}:=&\dfrac{\varepsilon_{\varepsilon_{ij}}}{2\varepsilon_{\varepsilon}}-\tfrac{1}{3}\delta_{ij}&&\quad;\quad
\II{b_{\varepsilon_\varepsilon}}=&&-\tfrac{1}{2}b_{\varepsilon_{\varepsilon_{mk}}}b_{\varepsilon_{\varepsilon_{km}}}&&\quad;\quad
\III{b_{\varepsilon_\varepsilon}}=&&\tfrac{1}{3}b_{\varepsilon_{\varepsilon_{mk}}}b_{\varepsilon_{\varepsilon_{k\ell}}}b_{\varepsilon_{\varepsilon_{\ell m}}}
                                                                                                                                    \label{Eq_FAepsijBsTPCF_s_epsepsij_002c}
\end{alignat}
lie within \possessivecite{Lumley_1978a} realisability triangle in the $(\III{},-\II{})$-plane \cite{Gerolymos_Vallet_2016b}.
\possessivecite{Lumley_1978a} flatness parameters
\begin{align}
A:=1+27\III{b}+9\II{b}\quad;\quad
A_\varepsilon:=1+27\III{b_\varepsilon}+9\II{b_\varepsilon}\quad;\quad
A_{\varepsilon_\varepsilon}:=1+27\III{b_{\varepsilon_\varepsilon}}+9\II{b_{\varepsilon_\varepsilon}}
                                                                                                                                    \label{Eq_FAepsijBsTPCF_s_epsepsij_002d}
\end{align}
\end{subequations}
are bounded in the interval $[0,1]$ \cite{Lumley_1978a}, between the 2-component (2-C) limit corresponding to the value $0$ and the isotropic componentality
corresponding to the value $1$ \cite{Simonsen_Krogstad_2005a}.
It is well known \cite{Mansour_Kim_Moin_1988a} that at the wall both $\tsr{r}$ and $\tsr{\varepsilon}$ reach the 2-C limit at the wall. It was recently shown \cite{Gerolymos_Vallet_2016b} that the 2-C limit at the wall
is approached quadratically ($A_\varepsilon\sim_{y^+\to0}4A\sim_{y^+\to0}O({y^+}^2)$). This result was obtained by calculating the wall-asymptotic expansions
of $\tsr{b}$ \cite[Tab. 1, p. 392]{Gerolymos_Vallet_2016b} and of $\tsr{b_{\varepsilon}}$ \cite[Tab. 2, p. 393]{Gerolymos_Vallet_2016b} and of their invariants.
However, as shown previously \figref{Fig_FAepsijBsTPCF_s_TPCFBs_ss_epsijvsrijBs_003} $\tsr{\varepsilon_{\varepsilon}}$ is not 2-C at the wall,
where all of its components are generally $\neq0$ \tabrefsab{Tab_FAepsijBsTPCF_s_WAs_002}
                                                            {Tab_FAepsijBsTPCF_s_epsepsij_001}.
\begin{figure}
\begin{center}
\begin{picture}(450,300)
\put(0,-15){\includegraphics[angle=0,width=370pt,bb=63 178 517 559]{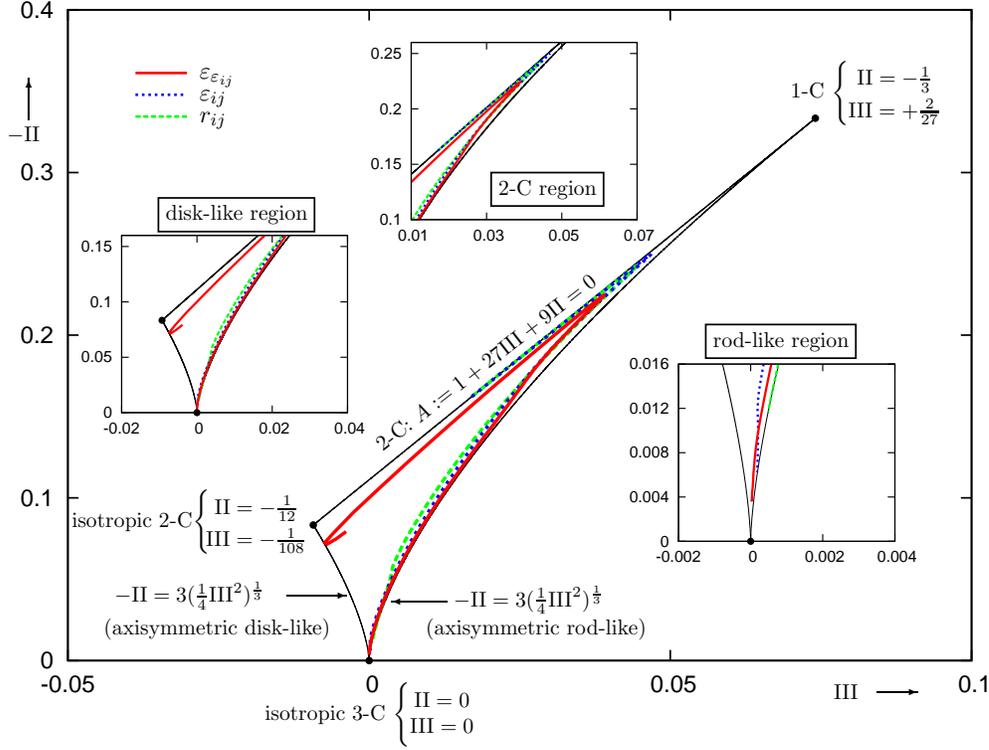}}
\end{picture}
\end{center}
\caption{\possessivecite{Lumley_1978a} realizability triangle \cite{Simonsen_Krogstad_2005a} in the $(-\mathrm{II},\mathrm{III})$-plane \eqref{Eq_FAepsijBsTPCF_s_epsepsij_002}
and trajectory in the wall-normal direction of the locus of the anisotropy-tensor invariants of the Reynolds-stresses $b_{ij}$ \eqref{Eq_FAepsijBsTPCF_s_epsepsij_002a},
of the dissipation tensor $b_{\varepsilon_{ij}}$ \eqref{Eq_FAepsijBsTPCF_s_epsepsij_002b} and of the destruction-of-dissipation tensor $b_{\varepsilon_{\varepsilon_{ij}}}$ \eqref{Eq_FAepsijBsTPCF_s_epsepsij_002c},
from the present \tsn{DNS} computations of turbulent plane channel flow ($Re_{\tau_w}\approxeq180$).}
\label{Fig_FAepsijBsTPCF_s_epsepsij_002}
\end{figure}

These differences in behaviour are better understood by considering \figref{Fig_FAepsijBsTPCF_s_epsepsij_001} the anisotropy tensors $\{\tsr{b},\tsr{b_{\varepsilon}},\tsr{b_{\varepsilon_{\varepsilon}}}\}$ 
and their invariants \eqref{Eq_FAepsijBsTPCF_s_epsepsij_002}.
Although the shear components $\{r_{xy}^+,\varepsilon_{xy}^+,\varepsilon_{\varepsilon_{xy}}^+\}$ are invariably much smaller than the traces $\{ {\rm k}^+,\varepsilon^+,\varepsilon_{\varepsilon}^+\}$
\eqref{Eq_FAepsijBsTPCF_s_epsepsij_001}, their anisotropy \figref{Fig_FAepsijBsTPCF_s_epsepsij_001} highlights some fundamental differences between the 3 tensors.
The shear Reynolds-stress is $r_{xy}^+<0\;\forall\;y^+\in\;]0,\delta^+[$ ($\mathrm{sign}\;r_{xy}=\mathrm{sign}\;b_{xy}$; \figrefnp{Fig_FAepsijBsTPCF_s_epsepsij_001}), 
whereas $\varepsilon_{xy}^+<0\;\forall\;y^+\in\;]0,\delta^+[$ is close to 0 at $y^+\approxeq 25$ 
\figref{Fig_FAepsijBsTPCF_s_epsepsij_001}, $\varepsilon_{\varepsilon_{xy}}^+$ exhibiting a radically different behaviour \figrefsab{Fig_FAepsijBsTPCF_s_TPCFBs_ss_epsijvsrijBs_003}
                                                                                                                                     {Fig_FAepsijBsTPCF_s_epsepsij_001}.
The wall-asymptotic expansion of $\tsr{b_{\varepsilon_{\varepsilon}}}$ \tabref{Tab_FAepsijBsTPCF_s_epsepsij_002} confirms that $\tsr{\varepsilon_{\varepsilon}}$ is not 2-C at the wall,
contrary to $\tsr{b}$ \cite[Tab. 1, p. 392]{Gerolymos_Vallet_2016b}
and $\tsr{b_{\varepsilon}}$ \cite[Tab. 2, p. 393]{Gerolymos_Vallet_2016b}.
This is clearly shown by the $y^+$-distribution of the corresponding flatness parameter \eqref{Eq_FAepsijBsTPCF_s_epsepsij_002d} $A_{\varepsilon_\varepsilon}>0\;\forall\;y^+$ \figref{Fig_FAepsijBsTPCF_s_epsepsij_001},
which reaches its minimum value $\approxeq 0.03$ at $y^+\approxeq 5$, then increasing to $[A_{\varepsilon_\varepsilon}]_w\approxeq 0.185$.
These differences in near-wall behaviour are also particularly visible in the $y^+$-distribution of the anisotropy invariants \figref{Fig_FAepsijBsTPCF_s_epsepsij_001}
and in the anisotropy invariant mapping (\tsn{AIM}) of $\tsr{\varepsilon_{\varepsilon}}$ \figref{Fig_FAepsijBsTPCF_s_epsepsij_002}. The locus of $\{\III{b_\varepsilon},-\II{b_\varepsilon}\}$ does not reach the 
2-C boundary \figref{Fig_FAepsijBsTPCF_s_epsepsij_002}. Instead, near the wall, $\{\III{b_\varepsilon},-\II{b_\varepsilon}\}$ reaches the axisymmetric disk-like boundary
of \possessivecite{Lumley_1978a} realisability triangle \figref{Fig_FAepsijBsTPCF_s_epsepsij_002},
roughly corresponding to $y^+\approxeq 0.7$ where $\varepsilon_{\varepsilon_{xx}}^+=\varepsilon_{\varepsilon_{zz}}^+$ \figref{Fig_FAepsijBsTPCF_s_TPCFBs_ss_epsijvsrijBs_003} 
and $b_{\varepsilon_{\varepsilon_{xx}}}=b_{\varepsilon_{\varepsilon_{zz}}}$ \figref{Fig_FAepsijBsTPCF_s_epsepsij_001},
also marked by the near-wall minimum of $\III{b_\varepsilon}$ \figref{Fig_FAepsijBsTPCF_s_epsepsij_001}. 
For $y^+\lessapprox 0.7$, the locus of $\tsr{b_{\varepsilon_{\varepsilon}}}$ in the $(\III{b_{\varepsilon_{\varepsilon}}},-\II{b_{\varepsilon_{\varepsilon}}})$-plane returns toward the interior of
\possessivecite{Lumley_1978a} realisability triangle \figref{Fig_FAepsijBsTPCF_s_epsepsij_002}.
The contrasting behaviour of $\tsr{b_{\varepsilon_{\varepsilon}}}$ compared to $\tsr{b}$ and $\tsr{b_\varepsilon}$ \figrefsab{Fig_FAepsijBsTPCF_s_epsepsij_001}
                                                                                                                             {Fig_FAepsijBsTPCF_s_epsepsij_002}
further highlights the complexity of near-wall turbulence, where 2-C componentality at the wall applies to both $\tsr{r}$ and $\tsr{\varepsilon}$ but not to $\tsr{\varepsilon_\varepsilon}$.
Examination of the wall-asymptotic behaviour of various terms in the $\varepsilon_{ij}$-budgets
\tabrefsab{Tab_FAepsijBsTPCF_s_WAs_001}
          {Tab_FAepsijBsTPCF_s_WAs_002}
reveals that neither $d_{\varepsilon_{ij}}^{(\mu)}$ nor $\Pi_{\varepsilon_{ij}}$ are 2-C at the wall, in line 
with \eqref{Eq_DTWT_s_epsijBs_ss_epsijTWAs_001b}, whereas $P_{\varepsilon_{ij}}$ and $d_{\varepsilon_{ij}}^{(u)}$ are 2-C at the wall \tabrefsab{Tab_FAepsijBsTPCF_s_WAs_001}
                                                                                                                                                {Tab_FAepsijBsTPCF_s_WAs_002}.
Notice in particular the wall-behaviour of $\Pi_{\varepsilon_{ij}}$, for which $[\Pi_{\varepsilon_{xy}}]_w^+\neq 0$ while $[\Pi_{\varepsilon_{yy}}]_w^+=0$ \tabref{Tab_FAepsijBsTPCF_s_WAs_001}.
Notice also that, at the wall, $\varepsilon_{\varepsilon_w}^{-1}\varepsilon_w$ defines, by dimensional analysis \cite[p. 5]{Tennekes_Lumley_1972a}, a time-scale which is finite contrary to ${\rm k}_w\varepsilon_w^{-1}=0$.

%
%
%
%
%
%
%
%
%
\section{Conclusions}\label{FAepsijBsTPCF_s_C}
%
%
%
%
%
%
%
%
%

The paper studies $\varepsilon_{ij}$-budgets, including the shear component, and compares the behaviour of different mechanisms with the corresponding mechanisms in
$r_{ij}$-budgets, using novel \tsn{DNS} data for low $Re_{\tau_w}\approxeq180$ plane channel flow.

All of the components of production $P_{\varepsilon_{ij}}$ are generally $\neq0$ (specifically all of the components of $P^{(2)}_{\varepsilon_{ij}}$ and $P^{(4)}_{\varepsilon_{ij}}$)
and contribute as gain to the corresponding $\varepsilon_{ij}$-budgets, contrary to the $r_{ij}$-budgets where for plane channel flow $P_{yy}=P_{zz}=0\;\forall\;y^+$.
The pressure mechanism $\Pi_{\varepsilon_{ij}}$ has a very weak contribution to the budgets of the streamwise $\varepsilon_{xx}$ and spanwise $\varepsilon_{zz}$ components,
in contrast to $\Pi_{ij}$ which is important in the budgets of all $r_{ij}$-components, especially in the log-region. The destruction-of-dissipation tensor $\varepsilon_{\varepsilon_{ij}}$
behaves very differently from the dissipation tensor $\varepsilon_{ij}$. The shear component $\varepsilon_{\varepsilon_{xy}}>0\;\forall\;y^+\gtrapprox3$ is a gain mechanism in the
$\varepsilon_{xy}$-budgets except very near the wall ($y^+\lessapprox3$), contrary to $\varepsilon_{xy}<0\;\forall\;y^+\in\;]0,\delta^+[$ which is a loss mechanism in the $r_{xy}$-budgets.
Finally, analytical results and \tsn{DNS} data for the wall-asymptotic behaviour of different terms in the $\varepsilon_{ij}$-budgets show that the
wall boundary-condition is $[d^{(\mu)}_{\varepsilon_{ij}}]^+_w+[\Pi_{\varepsilon_{ij}}]^+_w=[\varepsilon_{\varepsilon_{ij}}]^+_w$ instead of
the well known condition $[d^{(\mu)}_{ij}]^+_w=[\varepsilon_{ij}]^+_w$ for the $r_{ij}$-budgets \cite{Mansour_Kim_Moin_1988a}.

All of the 3 tensors ($r_{ij}$, $\varepsilon_{ij}$ and $\varepsilon_{\varepsilon_{ij}}$) being positive-definite, their anisotropy was studied using \tsn{AIM} \cite{Lee_Reynolds_1987a},
revealing in particular that, near the wall, the destruction-of-dissipation tensor $\varepsilon_{\varepsilon_{ij}}$,
after reaching the axisymmetric disk-like boundary (roughly where $\varepsilon_{\varepsilon_{xx}}\approxeq\varepsilon_{\varepsilon_{zz}}$,
at $y^+\approxeq0.7$), returns inside the realisability triangle,
never approaching the 2-C boundary. The \tsn{DNS} data are corroborated by the wall-asymptotic expansions of $\varepsilon_{\varepsilon_{ij}}$ and of its
anisotropy tensor $b_{\varepsilon_{\varepsilon_{ij}}}$. This observed componentality of $\varepsilon_{\varepsilon_{ij}}$ is strickingly different from that of $r_{ij}$ or $\varepsilon_{ij}$, both of which
are 2-C at the wall, and highlights the difference between componentality of various tensors and dimensionality of turbulence \cite{Kassinos_Reynolds_Rogers_2001a}.

The analysis of the \tsn{DNS} data highlights the complexity of $\varepsilon_{ij}$-transport, especially near the wall and regarding the shear component $\varepsilon_{xy}$.
It seems plausible that the specific behaviour of the $\varepsilon_{xy}$-budgets, both with respect to $r_{xy}$-budgets and compared to the diagonal components of $\varepsilon_{ij}$,
can only be modelled by differential $r_{ij}$--$\varepsilon_{ij}$ closures. It is hoped that the present \tsn{DNS} data will be useful in the development of such closures.

%
%
%
%
%
%
%
%
%
\section*{Acknowledgments}
%
%
%
%
%
%
%
%
%

The authors are listed alphabetically.
The computations reported in the present work were performed using \tsn{HPC} ressources allocated at \tsn{GENCI--IDRIS} (Grant 2015--022139)
and at \tsn{ICS--UPMC} (\tsn{ANR--10--EQPX--29--01}).
Tabulated \tsn{DNS} data are available at {\tt http://www.aerodynamics.fr/DNS\_database/CT\_chnnl}.
The present work was partly supported by the \tsn{ANR} project \NumERICCS (\tsn{ANR--15--CE06--0009}).

%
%
%
%
%
%
%
%
%
\appendix\section{Fully developed plane channel flow}\label{FAepsijBsTPCF_s_AppendixFDPCF}\renewcommand{\thesubsection}{\Alph{section}.\arabic{subsection}}\renewcommand{\thesubsubsection}{\Alph{section}.\arabic{subsection}.\arabic{subsubsection}}
%
%
%
%
%
%
%
%
%

We consider fully developed ($xz$-invariant) plane channel flow (the channel height is $2\delta$ and $xyz$ are respectively the streamwise, wall-normal and spanwise directions)
and use nondimensional inner variables \cite[wall-units, (A.3), p. 414]{Gerolymos_Vallet_2016b}.

%
%
%
%
%
\renewcommand{\thesubsection}{\Alph{section}.\arabic{subsection}}\subsection{Mean-flow and symmetries}\label{FAepsijBsTPCF_s_AppendixFDPCF_ss_MFS}
%
%
%
%
%

No-slip boundary-conditions apply at the walls
\begin{subequations}
                                                                                                                                    \label{Eq_FAepsijBsTPCF_s_AppendixFDPCF_ss_MFS_001}
\begin{alignat}{6}
&y^+\in\{0,2\delta^+\}\;\implies\;\bar u^+=\bar v^+=\bar w^+=u'^+=v'^+=w'^+=0\;\; ;\;\; \forall x^+,z^+,t^+
                                                                                                                                    \label{Eq_FAepsijBsTPCF_s_AppendixFDPCF_ss_MFS_001a}
\end{alignat}
The usual hypotheses that the mean-flow is steady, 2-D and that the $x$-wise location that is investigated is sufficiently downstream of the channel inlet to achieve
fully developed flow \cite{Zanoun_Nagib_Durst_2009a,Schultz_Flack_2013a}
in the streamwise direction
\begin{alignat}{6}
\dfrac{\partial \overline{(\cdot)}}
      {\partial t^+}=0\;\; ;\;\; 
\bar w^+ =0\;\; ;\;\; 
r_{yz}^+=r_{zx}^+=0\;\; ;\;\; 
\dfrac{\partial \overline{(\cdot)}}
       {\partial z^+}=0\;\; ;\;\; 
\dfrac{\partial \bar u_i^+}
      {\partial x^+}=0\;\; ;\;\; 
\dfrac{\partial r_{ij}^+}
      {\partial x^+}=0
                                                                                                                                    \label{Eq_FAepsijBsTPCF_s_AppendixFDPCF_ss_MFS_001b}
\end{alignat}
\end{subequations}
are made. Under these conditions \eqref{Eq_FAepsijBsTPCF_s_AppendixFDPCF_ss_MFS_001},
the mean continuity \cite[(4.5), p. 76]{Mathieu_Scott_2000a} and momentum (streamwise and wall-normal) equations \cite[(4.9), p. 77]{Mathieu_Scott_2000a}, imply \cite[pp. 105--111]{Mathieu_Scott_2000a} the exact relations
\begin{subequations}
                                                                                                                                    \label{Eq_FAepsijBsTPCF_s_AppendixFDPCF_ss_MFS_002}
\begin{alignat}{6}
\bar v^+ =& 0\quad\forall x^+,y^+
                                                                                                                                    \label{Eq_FAepsijBsTPCF_s_AppendixFDPCF_ss_MFS_002a}\\
\dfrac{\partial \bar p^+}
      {\partial      x^+}=&
\dfrac{d\bar p_w^+}
      {dx^+       }=-\left[\dfrac{\tau_w}{\delta}\right]^+=-\dfrac{1}{\delta^+}=-\dfrac{1}{Re_{\tau_w}}=\const\quad\forall x^+,y^+
                                                                                                                                    \label{Eq_FAepsijBsTPCF_s_AppendixFDPCF_ss_MFS_002b}\\
-r_{xy}^++\dfrac{d\bar u^+}{dy^+}=&\left(1-\dfrac{y^+}{Re_{\tau_w}}\right)
                                                                                                                                    \label{Eq_FAepsijBsTPCF_s_AppendixFDPCF_ss_MFS_002c}\\
\bar p^+(x^+,y^+)=&\;\bar p_w^+(x^+)-r_{yy}^+(y^+)
                                                                                                                                    \label{Eq_FAepsijBsTPCF_s_AppendixFDPCF_ss_MFS_002d}
\end{alignat}
for the mean streamwise velocity $\bar u^+(y^+)$ and mean pressure $\bar p^+(x^+,y^+)$ fields, with a constant streamwise pressure-gradient $\partial_x\bar p=d_x\bar p_w=\const$ \eqref{Eq_FAepsijBsTPCF_s_AppendixFDPCF_ss_MFS_002b}.
In \eqrefsab{Eq_FAepsijBsTPCF_s_AppendixFDPCF_ss_MFS_002b}
            {Eq_FAepsijBsTPCF_s_AppendixFDPCF_ss_MFS_002c}
$Re_{\tau_w}=\delta^+$ is the friction Reynolds number \cite[(A.3g), p. 414]{Gerolymos_Vallet_2016b}.
\end{subequations}
In \eqrefsab{Eq_FAepsijBsTPCF_s_AppendixFDPCF_ss_MFS_001}
            {Eq_FAepsijBsTPCF_s_AppendixFDPCF_ss_MFS_002}
deterministic potential body-forces (\eg\ gravity) in the momentum equations are included in the mean-pressure field \cite[p. 31]{Monin_Yaglom_1971a}.
Recall that the $xzt$-homogeneity of the averages implies the relations
\begin{subequations}
                                                                                                                                    \label{Eq_FAepsijBsTPCF_s_AppendixFDPCF_ss_MFS_003}
\begin{alignat}{6}
\dfrac{\partial\overline{(\cdot)'[\cdot]'}}{\partial q}=0\implies&
\overline{(\cdot)'\dfrac{\partial[\cdot]'}{\partial q}}=-\overline{[\cdot]'\dfrac{\partial(\cdot)'}{\partial q}}\quad\forall q\in\{x,z,t\}
                                                                                                                                    \label{Eq_FAepsijBsTPCF_s_AppendixFDPCF_ss_MFS_003a}\\
                                        &
\overline{(\cdot)'\dfrac{\partial^2[\cdot]'}{\partial q_1\partial q_2}}=-\overline{\dfrac{\partial(\cdot)'}{\partial q_1}\dfrac{\partial[\cdot]'}{\partial q_2}}
                                                                       =-\overline{\dfrac{\partial[\cdot]'}{\partial q_1}\dfrac{\partial(\cdot)'}{\partial q_2}}\quad\forall q_1,q_2\in\{x,z,t\}
                                                                                                                                    \label{Eq_FAepsijBsTPCF_s_AppendixFDPCF_ss_MFS_003b}
\end{alignat}
\end{subequations}

%
%
%
%
%
\subsection{$\varepsilon_{ij}$-budgets in plane channel flow}\label{FAepsijBsTPCF_s_AppendixFDPCF_ss_epsijBsPCF}
%
%
%
%
%

Under fully developed plane channel flow conditions \eqrefsab{Eq_FAepsijBsTPCF_s_AppendixFDPCF_ss_MFS_001}
                                                             {Eq_FAepsijBsTPCF_s_AppendixFDPCF_ss_MFS_002}
the $\varepsilon_{ij}$-transport equations simplify to
\begin{alignat}{6}
\eqrefsabc{Eq_FAepsijBsTPCF_s_TEqsWAs_ss_TEqs_001b}
          {Eq_FAepsijBsTPCF_s_AppendixFDPCF_ss_MFS_001}
          {Eq_FAepsijBsTPCF_s_AppendixFDPCF_ss_MFS_002}\implies&
\underbrace{\dfrac{d }
                  {dy}\left[-\rho\left(\overline{v'2\nu\dfrac{\partial u'_i}
                                                             {\partial  x_k}\dfrac{\partial u'_j}
                                                                                  {\partial  x_k}}\right)\right]}_{\displaystyle d_{\varepsilon_{ij}}^{(u)}}
+\underbrace{\mu\dfrac{d^2\varepsilon_{ij}}
                      {dy^2}}_{\displaystyle d_{\varepsilon_{ij}}^{(\mu)}}
\underbrace{-\rho\left(\varepsilon_{iy}\delta_{jx}+\varepsilon_{jy}\delta_{ix}\right)\dfrac{d\bar u}
                                                                                           {dy     }}_{\displaystyle P_{\varepsilon_{xx}}^{(1)}}
                                                                                                                                    \notag\\
\underbrace{-\rho\left(\mathcal{E}_{ijxy}+\mathcal{E}_{ijyx}\right)\dfrac{d\bar u}
                                                                         {dy     }}_{\displaystyle P_{\varepsilon_{ij}}^{(2)}}
&
\underbrace{-\rho\left(2\nu\overline{v'\dfrac{\partial u'_i}
                                             {\partial    y}}\delta_{jx}
                      +2\nu\overline{v'\dfrac{\partial u'_j}
                                             {\partial    y}}\delta_{ix}\right)\dfrac{d^2\bar u}
                                                                                     {dy^2     }}_{\displaystyle P_{\varepsilon_{ij}}^{(3)}}
+P_{\varepsilon_{ij}}^{(4)}+\Pi_{\varepsilon_{ij}}-\rho\varepsilon_{\varepsilon_{ij}}=0
                                                                                                                                    \label{Eq_FAepsijBsTPCF_s_AppendixFDPCF_ss_epsijBsPCF_001}
\end{alignat}
where $\mathcal{E}_{ijkm}:=2\nu\overline{\partial_{x_k}u_i'\partial_{x_m}u_j'}$ \cite[(3.1a), p. 402]{Gerolymos_Vallet_2016b} and
the 3 last terms in \eqref{Eq_FAepsijBsTPCF_s_AppendixFDPCF_ss_epsijBsPCF_001} retain their general expressions \eqref{Eq_FAepsijBsTPCF_s_TEqsWAs_ss_TEqs_001b}.
The relevant equations for the $\varepsilon_{ij}$-components (recall that by 2-D $z$-wise symmetry $\varepsilon^+_{yz}=\varepsilon^+_{zx}=0\;\forall y^+$) read in wall-units
\begin{subequations}
                                                                                                                                    \label{Eq_FAepsijBsTPCF_s_AppendixFDPCF_ss_epsijBsPCF_002}
\begin{alignat}{6}
\underbrace{\dfrac{d }
                  {dy}\left[-\rho\left(\overline{v'2\nu\dfrac{\partial  u'}
                                                             {\partial x_k}\dfrac{\partial  u'}
                                                                                 {\partial x_k}}\right)\right]}_{\displaystyle d_{\varepsilon_{xx}}^{(u)}}
+&\underbrace{\mu\dfrac{d^2\varepsilon_{xx}}
                      {dy^2}}_{\displaystyle d_{\varepsilon_{xx}}^{(\mu)}}
\underbrace{-2\rho\varepsilon_{xy}\dfrac{d\bar u}
                                        {dy     }}_{\displaystyle P_{\varepsilon_{xx}}^{(1)}}
\underbrace{-2\rho\mathcal{E}_{xxxy}\dfrac{d\bar u}
                                          {dy     }}_{\displaystyle P_{\varepsilon_{xx}}^{(2)}}
\underbrace{-4\rho\nu\overline{v'\dfrac{\partial u'}
                                       {\partial  y}}\dfrac{d^2\bar u}
                                                           {dy^2     }}_{\displaystyle P_{\varepsilon_{xx}}^{(3)}}
                                                                                                                                    \notag\\
+&P_{\varepsilon_{xx}}^{(4)}+\Pi_{\varepsilon_{xx}}-\rho\varepsilon_{\varepsilon_{xx}}\stackrel{\eqref{Eq_FAepsijBsTPCF_s_AppendixFDPCF_ss_epsijBsPCF_001}}{=}0
                                                                                                                                    \label{Eq_FAepsijBsTPCF_s_AppendixFDPCF_ss_epsijBsPCF_002a}
\end{alignat}
\begin{alignat}{6}
\underbrace{\dfrac{d }
                  {dy}\left[-\rho\left(\overline{v'2\nu\dfrac{\partial  u'}
                                                             {\partial x_k}\dfrac{\partial    v'}
                                                                                  {\partial  x_k}}\right)\right]}_{\displaystyle d_{\varepsilon_{xy}}^{(u)}}
+&\underbrace{\mu\dfrac{d^2\varepsilon_{xy}}
                      {dy^2}}_{\displaystyle d_{\varepsilon_{xy}}^{(\mu)}}
\underbrace{-\rho\varepsilon_{yy}\dfrac{d\bar u}
                                       {dy     }}_{\displaystyle P_{\varepsilon_{xy}}^{(1)}}
\underbrace{-\rho\left(\mathcal{E}_{xyxy}+\mathcal{E}_{xyyx}\right)\dfrac{d\bar u}
                                                                         {dy     }}_{\displaystyle P_{\varepsilon_{xy}}^{(2)}}
\underbrace{-\rho2\nu\overline{v'\dfrac{\partial v'}
                                       {\partial  y}}\dfrac{d^2\bar u}
                                                           {dy^2   }}_{\displaystyle P_{\varepsilon_{xy}}^{(3)}}
                                                                                                                                    \notag\\
+&P_{\varepsilon_{xy}}^{(4)}+\Pi_{\varepsilon_{xy}}-\rho\varepsilon_{\varepsilon_{xy}}\stackrel{\eqref{Eq_FAepsijBsTPCF_s_AppendixFDPCF_ss_epsijBsPCF_001}}{=}0
                                                                                                                                    \label{Eq_FAepsijBsTPCF_s_AppendixFDPCF_ss_epsijBsPCF_002b}
\end{alignat}
\begin{alignat}{6}
\underbrace{\dfrac{d }
                  {dy}\left[-\rho\left(\overline{v'2\nu\dfrac{\partial  v'}
                                                             {\partial x_k}\dfrac{\partial  v'}
                                                                                 {\partial x_k}}\right)\right]}_{\displaystyle d_{\varepsilon_{yy}}^{(u)}}
+&\underbrace{\mu\dfrac{d^2\varepsilon_{yy}}
                      {dy^2}}_{\displaystyle d_{\varepsilon_{yy}}^{(\mu)}}
+\underbrace{0}_{\displaystyle P_{\varepsilon_{yy}}^{(1)}}
\underbrace{-2\rho\mathcal{E}_{yyxy}\dfrac{d\bar u}
                                          {dy     }}_{\displaystyle P_{\varepsilon_{yy}}^{(2)}}
+\underbrace{0}_{\displaystyle P_{\varepsilon_{yy}}^{(3)}}
                                                                                                                                    \notag\\
+&P_{\varepsilon_{yy}}^{(4)}+\Pi_{\varepsilon_{yy}}-\rho\varepsilon_{\varepsilon_{yy}}\stackrel{\eqref{Eq_FAepsijBsTPCF_s_AppendixFDPCF_ss_epsijBsPCF_001}}{=}0
                                                                                                                                    \label{Eq_FAepsijBsTPCF_s_AppendixFDPCF_ss_epsijBsPCF_002c}
\end{alignat}
\begin{alignat}{6}
\underbrace{\dfrac{d }
                  {dy}\left[-\rho\left(\overline{v'2\nu\dfrac{\partial  w'}
                                                             {\partial x_k}\dfrac{\partial w' }
                                                                                 {\partial x_k}}\right)\right]}_{\displaystyle d_{\varepsilon_{zz}}^{(u)}}
+&\underbrace{\mu\dfrac{d^2\varepsilon_{zz}}
                      {dy^2}}_{\displaystyle d_{\varepsilon_{zz}}^{(\mu)}}
+\underbrace{0}_{\displaystyle P_{\varepsilon_{zz}}^{(1)}}
\underbrace{-2\rho\mathcal{E}_{zzxy}\dfrac{d\bar u}
                                          {dy     }}_{\displaystyle P_{\varepsilon_{zz}}^{(2)}}
+\underbrace{0}_{\displaystyle P_{\varepsilon_{zz}}^{(3)}}
                                                                                                                                    \notag\\
+&P_{\varepsilon_{zz}}^{(4)}+\Pi_{\varepsilon_{zz}}-\rho\varepsilon_{\varepsilon_{zz}}\stackrel{\eqref{Eq_FAepsijBsTPCF_s_AppendixFDPCF_ss_epsijBsPCF_001}}{=}0
                                                                                                                                    \label{Eq_FAepsijBsTPCF_s_AppendixFDPCF_ss_epsijBsPCF_002d}
\end{alignat}
\end{subequations}
where the symmetry relations $\mathcal{E}_{xxxy}=\mathcal{E}_{xxyx}$, $\mathcal{E}_{yyxy}=\mathcal{E}_{yyyx}$ and $\mathcal{E}_{zzxy}=\mathcal{E}_{zzyx}$ were used.

%
%
%
%
%
%
%
%
%
\section{Asymptotic behaviour in the viscous sublayer ($y^+\to0$)}\label{FAepsijBsTPCF_s_AppendixABVSy+0}
%
%
%
%
%
%
%
%
%

Near a plane $xz$-wall, located at $y^+=0$, the fluctuating quantities are expandend $y$-wise in Taylor-series around $y^+=0$ following \eqref{Eq_FAepsijBsTPCF_s_WAs_001}.
The application of the usual gradient-operator \cite[(A.48), p. 651]{Pope_2000a} $\nabla(\cdot):=\vec{e}_\ell\partial_{x_\ell}(\cdot)$ on the coefficients of \eqref{Eq_FAepsijBsTPCF_s_WAs_001},
which are stationary random functions of $\{x^+,z^+,t^+\}$ independent of $y^+$, produces only in-plane $xz$-gradients
\begin{subequations}
                                                                                                                                    \label{Eq_FAepsijBsTPCF_s_AppendixABVSy+0_001}
\begin{alignat}{6}
\Big(\nabla(\cdot)_w'\Big)^+   &=&\vec{e}_x\dfrac{\partial(\cdot)_w'^+}{\partial x^+}   &+&\vec{e}_z\dfrac{\partial(\cdot)_w'^+}{\partial z^+}
                                                                                                                                    \label{Eq_FAepsijBsTPCF_s_AppendixABVSy+0_001a}\\
\Big(\nabla A_{(\cdot)}'\Big)^+&=&\vec{e}_x\dfrac{\partial A_{(\cdot)}'^+}{\partial x^+}&+&\vec{e}_z\dfrac{\partial A_{(\cdot)}'^+}{\partial z^+}
                                                                                                                                    \label{Eq_FAepsijBsTPCF_s_AppendixABVSy+0_001b}\\
\Big(\nabla B_{(\cdot)}'\Big)^+&=&\vec{e}_x\dfrac{\partial B_{(\cdot)}'^+}{\partial x^+}&+&\vec{e}_z\dfrac{\partial B_{(\cdot)}'^+}{\partial z^+}
                                                                                                                                    \label{Eq_FAepsijBsTPCF_s_AppendixABVSy+0_001c}\\
                               &\vdots&
                                                                                                                                    \notag
\end{alignat}
\end{subequations}

%
%
%
%
%
\subsection{Fluctuating continuity equation}\label{FAepsijBsTPCF_s_AppendixABVSy+0_ss_FCEq}
%
%
%
%
%

The no-slip condition \eqref{Eq_FAepsijBsTPCF_s_AppendixFDPCF_ss_MFS_001a} implies that the wall-terms in the expansions \eqref{Eq_FAepsijBsTPCF_s_WAs_001}
\begin{alignat}{6}
u_w'^+=v_w'^+=w_w'^+=0\qquad\forall\;x^+,z^+,t^+
                                                                                                                                    \label{Eq_FAepsijBsTPCF_s_AppendixABVSy+0_ss_FCEq_001}
\end{alignat}
Using the expansions \eqref{Eq_FAepsijBsTPCF_s_WAs_001}, along with \eqref{Eq_FAepsijBsTPCF_s_AppendixABVSy+0_ss_FCEq_001},
in the fluctuating continuity equation \cite[(4.6), p. 76]{Mathieu_Scott_2000a}
\begin{alignat}{6}
                         \dfrac{\partial    u'^+_\ell}
                               {\partial     x^+_\ell}=0
                                                                                                                                    \label{Eq_FAepsijBsTPCF_s_AppendixABVSy+0_ss_FCEq_002}
\end{alignat}
and equating the coefficients of different powers of $y^+$ to $0$, yields
\begin{subequations}
                                                                                                                                    \label{Eq_FAepsijBsTPCF_s_AppendixABVSy+0_ss_FCEq_003}\\
\begin{alignat}{6}
                                                                              A_v'^+=&0
                                                                                                                                    \label{Eq_FAepsijBsTPCF_s_AppendixABVSy+0_ss_FCEq_003a}\\
\dfrac{\partial A_w'^+}{\partial z^+}+\dfrac{\partial A_u'^+}{\partial x^+}+2 B_v'^+=&0
                                                                                                                                    \label{Eq_FAepsijBsTPCF_s_AppendixABVSy+0_ss_FCEq_003b}\\
\dfrac{\partial B_w'^+}{\partial z^+}+\dfrac{\partial B_u'^+}{\partial x^+}+3 C_v'^+=&0
                                                                                                                                    \label{Eq_FAepsijBsTPCF_s_AppendixABVSy+0_ss_FCEq_003c}
\end{alignat}
\end{subequations}
respectively for the $\{O(1),O(y^+),O({y^+}^2)\}$ terms, with analogous relations for \tsn{HOT}s.
Relation \eqref{Eq_FAepsijBsTPCF_s_AppendixABVSy+0_ss_FCEq_003b} corresponds to \citeasnoun[(3), p. 19]{Mansour_Kim_Moin_1988a}.
Notice that \eqref{Eq_FAepsijBsTPCF_s_AppendixABVSy+0_ss_FCEq_003b} yields the identity
\begin{alignat}{6}
\overline{{B_v'^+}^2}=&-\tfrac{1}{2}\overline{B_v'^+\dfrac{\partial A_u'^+}{\partial x^+}}-\tfrac{1}{2}\overline{B_v'^+\dfrac{\partial A_w'^+}{\partial z^+}}
                                                                                                                                    \label{Eq_FAepsijBsTPCF_s_AppendixABVSy+0_ss_FCEq_004}
\end{alignat}
Relations \eqrefsabcd{Eq_FAepsijBsTPCF_s_WAs_001}
                     {Eq_FAepsijBsTPCF_s_AppendixABVSy+0_ss_FCEq_001}
                     {Eq_FAepsijBsTPCF_s_AppendixABVSy+0_ss_FCEq_003}
                     {Eq_FAepsijBsTPCF_s_AppendixABVSy+0_ss_FCEq_004}
are generally valid for $xz$-inhomogeneous incompressible flow near an $xz$-wall. They provide the wall-asymptotic expansions of all correlations containing only fluctuating velocities and their derivatives,
and were used to calculate the wall-asymptotic expansions of $\varepsilon_{\varepsilon_{ij}}$ \tabref{Tab_FAepsijBsTPCF_s_epsepsij_001}
and of its anisotropy tensor $b_{\varepsilon_{\varepsilon_{ij}}}$ and invariants \tabref{Tab_FAepsijBsTPCF_s_epsepsij_002}.
The relation of the wall-asymptotic expansion of the fluctuating pressure $p'$ to the expansions of the fluctuating velocities depends on the particular mean-flow studied, and was therefore calculated
for fully developed plane channel flow.

%
%
%
%
%
\subsection{Plane channel flow}\label{FAepsijBsTPCF_s_AppendixABVSy+0_ss_PCF}
%
%
%
%
%

In the particular case of plane channel flow, conditions \eqrefsatob{Eq_FAepsijBsTPCF_s_AppendixFDPCF_ss_MFS_001}
                                                                    {Eq_FAepsijBsTPCF_s_AppendixFDPCF_ss_MFS_003}
imply specific relations for the mean and fluctuating fields, which were used to determine the wall-asymptotic expansions \tabrefsab{Tab_FAepsijBsTPCF_s_WAs_001}
                                                                                                                                    {Tab_FAepsijBsTPCF_s_WAs_002}
of various terms in the $\varepsilon_{ij}$-transport \eqref{Eq_FAepsijBsTPCF_s_TEqsWAs_ss_TEqs_001b} simplified for
plane channel flow \eqrefsab{Eq_FAepsijBsTPCF_s_AppendixFDPCF_ss_epsijBsPCF_001}
                            {Eq_FAepsijBsTPCF_s_AppendixFDPCF_ss_epsijBsPCF_002}.

%
\subsubsection{Mean-flow}\label{DTWT_s_AppendixABVSy+0_ss_PCF_sss_MF}
%

Using the expansion of $r_{xy}^+$ obtained from \eqrefsabc{Eq_FAepsijBsTPCF_s_WAs_001}
                                                          {Eq_FAepsijBsTPCF_s_AppendixABVSy+0_ss_FCEq_001}
                                                          {Eq_FAepsijBsTPCF_s_AppendixABVSy+0_ss_FCEq_003a}
in the $x$-wise component of the mean-momentum equation \eqref{Eq_FAepsijBsTPCF_s_AppendixFDPCF_ss_MFS_002c} yields, after integration and application
of the no-slip boundary-condition \eqref{Eq_FAepsijBsTPCF_s_AppendixFDPCF_ss_MFS_001a}, the expansion of the mean streamwise velocity
\begin{subequations}
                                                                                                                                    \label{Eq_FAepsijBsTPCF_s_AppendixABVSy+0_ss_PCF_sss_MF_001}
\begin{alignat}{6}
\!\!\!\!
\bar u^+\!\!\underset{y^+\to0}{\sim} y^+\!-\dfrac{1}{2 Re_{\tau_w}}{y^+}^2\!+\tfrac{1}{4}\overline{A_u'^+B_v'^+}{y^+}^4\!+\tfrac{1}{5}\left(\overline{B_u'^+B_v'^+}\!+\!\overline{A_u'^+C_v'^+}\right){y^+}^5\!+O({y^+}^6)
                                                                                                                                    \label{Eq_FAepsijBsTPCF_s_AppendixABVSy+0_ss_PCF_sss_MF_001a}
\end{alignat}
including the dominant linear term $y^+$, an $O({y^+}^2)$ correction associated with the mean streamwise pressure-gradient \eqref{Eq_FAepsijBsTPCF_s_AppendixFDPCF_ss_MFS_002b},
which $\to0$ as $Re_{\tau_w}\to\infty$ at fixed $y^+$, and higher $O({y^+}^4)$ terms.
Therefore, the gradient $[d_y\bar u]^+$ and Hessian $[d^2_{yy}\bar u]^+$ which appear in the production terms $\{P_{\varepsilon_{ij}}^{(1)},P_{\varepsilon_{ij}}^{(2)},P_{\varepsilon_{ij}}^{(3)}\}$
of the $\varepsilon_{ij}$-transport equations \eqrefsabc{Eq_FAepsijBsTPCF_s_TEqsWAs_ss_TEqs_001b}
                                                        {Eq_FAepsijBsTPCF_s_AppendixFDPCF_ss_epsijBsPCF_001}
                                                        {Eq_FAepsijBsTPCF_s_AppendixFDPCF_ss_epsijBsPCF_002}
expand as
\begin{alignat}{8}
\dfrac{d\bar u^+}
      {     dy^+}\underset{y^+\to0}{\sim}  &1&-&\dfrac{1}{Re_{\tau_w}}&y^+&&+& \overline{A_u'^+B_v'^+}&{y^+}^3&&+& \left(\overline{B_u'^+B_v'^+}+\overline{A_u'^+C_v'^+}\right)&{y^+}^4&&+&O({y^+}^5)
                                                                                                                                    \label{Eq_FAepsijBsTPCF_s_AppendixABVSy+0_ss_PCF_sss_MF_001b}\\
\dfrac{d^2\bar u^+}
      {   d{y^+}^2}\underset{y^+\to0}{\sim}& &-&\dfrac{1}{Re_{\tau_w}}&   &&+&3\overline{A_u'^+B_v'^+}&{y^+}^2&&+&4\left(\overline{B_u'^+B_v'^+}+\overline{A_u'^+C_v'^+}\right)&{y^+}^3&&+&O({y^+}^4)
                                                                                                                                    \label{Eq_FAepsijBsTPCF_s_AppendixABVSy+0_ss_PCF_sss_MF_001c}
\end{alignat}
By \eqrefsabcd{Eq_FAepsijBsTPCF_s_AppendixFDPCF_ss_MFS_002d}
              {Eq_FAepsijBsTPCF_s_WAs_001}
              {Eq_FAepsijBsTPCF_s_AppendixABVSy+0_ss_FCEq_001}
              {Eq_FAepsijBsTPCF_s_AppendixABVSy+0_ss_FCEq_003a},
the mean pressure can be expanded as
\begin{alignat}{6}
\!\!\!\!
\bar p^+\!\!\underset{y^+\to0}{\sim} \bar p_w^+(x)-\overline{{B_v'^+}^2}{y^+}^4-2\overline{B_v'^+C_v'^+}{y^+}^5-(2\overline{B_v'^+D_v'^+}\!+\!\overline{{C_v'^+}^2}){y^+}^6+O({y^+}^7)
                                                                                                                                    \label{Eq_FAepsijBsTPCF_s_AppendixABVSy+0_ss_PCF_sss_MF_001d}
\end{alignat}
\end{subequations}

%
\subsubsection{Wall-normal ($y$) fluctuating momentum and fluctuating pressure field}\label{FAepsijBsTPCF_s_AppendixABVSy+0_ss_PCF_sss_WNFMFPF}
%

Using \eqrefsabcd{Eq_FAepsijBsTPCF_s_WAs_001}
                 {Eq_FAepsijBsTPCF_s_AppendixABVSy+0_ss_FCEq_001}
                 {Eq_FAepsijBsTPCF_s_AppendixABVSy+0_ss_FCEq_003a}
                 {Eq_FAepsijBsTPCF_s_AppendixABVSy+0_ss_PCF_sss_MF_001a}
in the wall-normal component of the fluctuating momentum equation \cite[(4.31), p. 85]{Mathieu_Scott_2000a}
\begin{alignat}{6}
\dfrac{\partial u'^+_i}
      {\partial    t^+}+\bar u^+_\ell\dfrac{\partial   u'^+_i}
                                           {\partial x^+_\ell} =-\dfrac{\partial       }
                                                                       {\partial x^+_\ell}\left(u'^+_iu'^+_\ell-r^+_{i\ell}\right)
                                                                   -u'^+_\ell\dfrac{\partial\bar u^+_i}
                                                                                   {\partial  x^+_\ell}-\dfrac{\partial  p'^+}
                                                                                                               {\partial x^+_i}+\dfrac{\partial^2                  u'^+_i}
                                                                                                                                      {\partial x^+_\ell\partial x^+_\ell}
                                                                                                                                    \label{Eq_FAepsijBsTPCF_s_AppendixABVSy+0_ss_PCF_sss_WNFMFPF_001}
\end{alignat}
and using the symmetry conditions \eqref{Eq_FAepsijBsTPCF_s_AppendixFDPCF_ss_MFS_001b}
implies that the fluctuating pressure field expansion \eqref{Eq_FAepsijBsTPCF_s_WAs_001} should be
\begin{alignat}{6}
\!\!\!\!
p'^+\!\!\underset{y^+\to0}{\sim} p_w'^++2B_v'^+y^++3C_v'^+{y^+}^2+\tfrac{1}{3}\left(12D_v'^++\left(\nabla^2B_v'\right)^+-\dfrac{\partial B_v'^+}{\partial t^+}\right){y^+}^3+\cdots
                                                                                                                                    \label{Eq_FAepsijBsTPCF_s_AppendixABVSy+0_ss_PCF_sss_WNFMFPF_002}
\end{alignat}
\ie\ that the fluctuating pressure field, as $y^+\to0$, is uniquely determined to $O({y^+}^3)$ by the wall-normal fluctuating velocity field $v'$ \cite[(2.3b), p. 391]{Gerolymos_Vallet_2016b},
in line with the plane wall boundary condition $\partial_y p'=\mu\partial^2_{yy}v'$ \cite[(11.173), p. 439]{Pope_2000a}.
Relation \eqref{Eq_FAepsijBsTPCF_s_AppendixABVSy+0_ss_PCF_sss_WNFMFPF_002} corresponds to \citeasnoun[(2, 6), pp. 18--20]{Mansour_Kim_Moin_1988a}.
In \eqref{Eq_FAepsijBsTPCF_s_AppendixABVSy+0_ss_PCF_sss_WNFMFPF_002} $p_w'^+(x^+,z^+,t^+)$ is the fluctuating pressure at the wall.

%
\subsubsection{Wall-parallel ($xz$) fluctuating momentum}\label{FAepsijBsTPCF_s_AppendixABVSy+0_ss_PCF_sss_WPFM}
%

Using the expansions \eqrefsabcd{Eq_FAepsijBsTPCF_s_WAs_001}
                                {Eq_FAepsijBsTPCF_s_AppendixABVSy+0_ss_FCEq_001}
                                {Eq_FAepsijBsTPCF_s_AppendixABVSy+0_ss_FCEq_003a}
                                {Eq_FAepsijBsTPCF_s_AppendixABVSy+0_ss_PCF_sss_MF_001}
in the fluctuating $x$-momentum equation \eqref{Eq_FAepsijBsTPCF_s_AppendixABVSy+0_ss_PCF_sss_WNFMFPF_001}, and equating the coefficients of different powers of $y^+$ to $0$, yields 
\begin{subequations}
                                                                                                                                    \label{Eq_FAepsijBsTPCF_s_AppendixABVSy+0_ss_PCF_sss_WPFM_001}
\begin{alignat}{6}
&\dfrac{\partial p_w'^+}{\partial x^+}-2 B_u'^+=0
                                                                                                                                    \label{Eq_FAepsijBsTPCF_s_AppendixABVSy+0_ss_PCF_sss_WPFM_001a}\\
&2\dfrac{\partial B_v'^+}{\partial x^+}-\dfrac{\partial^2 A_u'^+}{\partial {z^+}^2}-\dfrac{\partial^2 A_u'^+}{\partial {x^+}^2}+\dfrac{\partial A_u'^+}{\partial t^+}-6 C_u'^+=0
                                                                                                                                    \label{Eq_FAepsijBsTPCF_s_AppendixABVSy+0_ss_PCF_sss_WPFM_001b}
\end{alignat}
\end{subequations}
respectively for the $\{O(1),O(y^+)\}$ terms, with the corresponding relations
\begin{subequations}
                                                                                                                                    \label{Eq_FAepsijBsTPCF_s_AppendixABVSy+0_ss_PCF_sss_WPFM_002}
\begin{alignat}{6}
&\dfrac{\partial p_w'^+}{\partial z^+}-2 B_w'^+=0
                                                                                                                                    \label{Eq_FAepsijBsTPCF_s_AppendixABVSy+0_ss_PCF_sss_WPFM_002a}\\
&2\dfrac{\partial B_v'^+}{\partial z^+}-\dfrac{\partial^2 A_w'^+}{\partial {z^+}^2}-\dfrac{\partial^2 A_w'^+}{\partial {x^+}^2}+\dfrac{\partial A_w'^+}{\partial t^+}-6 C_w'^+=0
                                                                                                                                    \label{Eq_FAepsijBsTPCF_s_AppendixABVSy+0_ss_PCF_sss_WPFM_002b}
\end{alignat}
\end{subequations}
for the fluctuating $z$-momentum equation \eqref{Eq_FAepsijBsTPCF_s_AppendixABVSy+0_ss_PCF_sss_WNFMFPF_001}.
Relations \eqrefsab{Eq_FAepsijBsTPCF_s_AppendixABVSy+0_ss_PCF_sss_WPFM_001a}
                   {Eq_FAepsijBsTPCF_s_AppendixABVSy+0_ss_PCF_sss_WPFM_002a}
correspond to \citeasnoun[(4), p. 19]{Mansour_Kim_Moin_1988a}
and relations \eqrefsab{Eq_FAepsijBsTPCF_s_AppendixABVSy+0_ss_PCF_sss_WPFM_001b}
                       {Eq_FAepsijBsTPCF_s_AppendixABVSy+0_ss_PCF_sss_WPFM_002b}
to \citeasnoun[(7, 8), p. 20]{Mansour_Kim_Moin_1988a}.

By \eqrefsab{Eq_FAepsijBsTPCF_s_AppendixABVSy+0_ss_PCF_sss_WPFM_001a}
            {Eq_FAepsijBsTPCF_s_AppendixABVSy+0_ss_PCF_sss_WPFM_002a},
\begin{subequations}
                                                                                                                                    \label{Eq_FAepsijBsTPCF_s_AppendixABVSy+0_ss_PCF_sss_WPFM_003}
\begin{alignat}{6}
\eqrefsab{Eq_FAepsijBsTPCF_s_AppendixABVSy+0_ss_PCF_sss_WPFM_001a}
         {Eq_FAepsijBsTPCF_s_AppendixABVSy+0_ss_PCF_sss_WPFM_002a}\implies&\tfrac{1}{2}\dfrac{\partial^2p_w'^+}{\partial x^+\partial z^+}=\dfrac{\partial B_u'^+}{\partial z^+}=\dfrac{\partial B_w'^+}{\partial x^+}
                                                                                                                                    \label{Eq_FAepsijBsTPCF_s_AppendixABVSy+0_ss_PCF_sss_WPFM_003a}
\end{alignat}
whence, using \eqref{Eq_FAepsijBsTPCF_s_AppendixFDPCF_ss_MFS_003a},
\begin{alignat}{6}
\eqrefsab{Eq_FAepsijBsTPCF_s_AppendixABVSy+0_ss_PCF_sss_WPFM_003a}
         {Eq_FAepsijBsTPCF_s_AppendixFDPCF_ss_MFS_003a}\implies&\overline{B_u'^+\dfrac{\partial B_w'^+}{\partial x^+}}=
                                                                  \overline{B_w'^+\dfrac{\partial B_u'^+}{\partial z^+}}=
                                                                  \overline{B_w'^+\dfrac{\partial B_u'^+}{\partial x^+}}=
                                                                  \overline{B_u'^+\dfrac{\partial B_w'^+}{\partial z^+}}=0
                                                                                                                                    \label{Eq_FAepsijBsTPCF_s_AppendixABVSy+0_ss_PCF_sss_WPFM_003b}
\end{alignat}
Notice that the relations $\overline{B_u'^+\partial_{x^+}B_w'^+}\stackrel{\eqref{Eq_FAepsijBsTPCF_s_AppendixFDPCF_ss_MFS_003a}}{=}-\overline{B_w'^+\partial_{x^+}B_u'^+}$
is also obvious because the flow is 2-D $z$-wise.
Relation \eqref{Eq_FAepsijBsTPCF_s_AppendixABVSy+0_ss_PCF_sss_WPFM_003a} corresponds to \citeasnoun[(5), p. 19]{Mansour_Kim_Moin_1988a}.
Furthermore, substituting $C_v'^+$ by \eqref{Eq_FAepsijBsTPCF_s_AppendixABVSy+0_ss_FCEq_003c} in $\overline{B_u'^+C_v^+}$ readily yields by \eqrefsab{Eq_FAepsijBsTPCF_s_AppendixFDPCF_ss_MFS_003a}
                                                                                                                                                       {Eq_FAepsijBsTPCF_s_AppendixABVSy+0_ss_PCF_sss_WPFM_003b}
\begin{alignat}{6}
\overline{B_u'^+C_v'^+}\stackrel{\eqrefsabc{Eq_FAepsijBsTPCF_s_AppendixABVSy+0_ss_FCEq_003c}
                                           {Eq_FAepsijBsTPCF_s_AppendixFDPCF_ss_MFS_003a}
                                           {Eq_FAepsijBsTPCF_s_AppendixABVSy+0_ss_PCF_sss_WPFM_003b}}{=}0
                                                                                                                                    \label{Eq_FAepsijBsTPCF_s_AppendixABVSy+0_ss_PCF_sss_WPFM_003c}
\end{alignat}
the corresponding relation $\overline{B_w'^+C_v^+}=0$, which can also be proven in the same way from \eqref{Eq_FAepsijBsTPCF_s_AppendixABVSy+0_ss_PCF_sss_WPFM_003b},
being obvious because the flow is 2-D in the mean $z$-wise \eqref{Eq_FAepsijBsTPCF_s_AppendixFDPCF_ss_MFS_001b}.
\end{subequations}
Finally, by \eqrefsab{Eq_FAepsijBsTPCF_s_AppendixABVSy+0_ss_PCF_sss_WPFM_001b}
                     {Eq_FAepsijBsTPCF_s_AppendixFDPCF_ss_MFS_003a}
\begin{alignat}{6}
\overline{B_v'^+\dfrac{\partial A_u'^+}{\partial t^+}}
\stackrel{\eqrefsab{Eq_FAepsijBsTPCF_s_AppendixABVSy+0_ss_PCF_sss_WPFM_001b}
                   {Eq_FAepsijBsTPCF_s_AppendixFDPCF_ss_MFS_003a}}{=}\,6\,\overline{B_v'^+C_u'^+}-\overline{\dfrac{\partial A_u'^+}{\partial z^+}\dfrac{\partial B_v'^+}{\partial z^+}}- \overline{\dfrac{\partial A_u'^+}{\partial x^+}\dfrac{\partial B_v'^+}{\partial x^+}}
                                                                                                                                    \label{Eq_FAepsijBsTPCF_s_AppendixABVSy+0_ss_PCF_sss_WPFM_004}
\end{alignat}

%
%
%
%
%
%
%
%
%
\section*{References}\bibliographystyle{dcu}\footnotesize\bibliography{Aerodynamics,GV,GV_news}\normalsize
%
%
%
%
%
%
%
%
%

\end{document}

%% file: Tab_FluidDynRes_01_WAs_epsijT.tex
\scalefont{1.0}{
\begin{alignat}{6}
d_{\varepsilon_{xx}}^{(\mu)+}&\sim&&4\left(6   \overline{A_u'^+C_u'^+}
                                          +4   \overline{{B_u'^+}^2}      
                                          +    \overline{{(\nabla A_u')^+}^2}\right)  \notag\\
                             &    +&& 24\left(4\overline{A_u'^+D_u'^+}
                                           +6\overline{B_u'^+C_u'^+}
                                           + \overline{(\nabla A_u')^+\cdot(\nabla B_u')^+}\right){y^+}
                               +  O({y^+}^2)
                                                                                                                                    \notag\\
d_{\varepsilon_{xy}}^{(\mu)+}&\sim&&4\left(3\overline{A_u'^+C_v'^+}
                                          +4\overline{B_u'^+B_v'^+}\right )            \notag\\
                             &  + && 12\left(4\overline{A_u'^+D_v'^+}
                                         +6\overline{B_v'^+C_u'^+}
                                         + \overline{(\nabla A_u')^+\cdot(\nabla B_v')^+}\right) {y^+}
                               +O({y^+}^2)
                                                                                                                                    \notag\\
d_{\varepsilon_{yy}}^{(\mu)+}&\sim&&16\,\overline{{B_v'^+}^2}
                                 + 144\,\overline{B_v'^+C_v'^+}  {y^+}
                               +  O({y^+}^2)
                                                                                                                                    \notag\\
d_{\varepsilon_{zz}}^{(\mu)+}&\sim&&4\left (6\overline{A_w'^+C_w'^+}
                                          +4\overline{{B_w'^+}^2}
                                          + \overline{{(\nabla A_w')^+}^2}\right)  \notag\\
                             &   +&& 24\left( 4\overline{A_w'^+D_w'^+}           
                                          +6\overline{B_w'^+C_w'^+}
                                          + \overline{(\nabla A_w')^+\cdot(\nabla B_w')^+}\right) {y^+}
                               +  O({y^+}^2)
                                                                                                                                    \notag
\end{alignat}
\vspace{-0.25in}
\begin{alignat}{6}
d_{\varepsilon_{xx}}^{(u)+}\sim&-4\overline{{A_u'^+}^2B_v'^+}{y^+}-6\left(\overline{{A_u'^+}^2C_v'^+}+4\overline{A_u'^+B_u'^+B_v'^+}\right){y^+}^2+O({y^+}^3)
                                                                                                                                    \notag\\
d_{\varepsilon_{xy}}^{(u)+}\sim&-12\overline{A_u'^+{B_v'^+}^2}{y^+}^2-\left(40\overline{A_u'^+B_v'^+C_v'^+}+32\overline{B_u'^+{B_v'^+}^2}\right){y^+}^3+O({y^+}^4)
                                                                                                                                    \notag\\
d_{\varepsilon_{yy}}^{(u)+}\sim&-32\overline{{B_v'^+}^3} {y^+}^3-160\overline{{B_v'^+}^2C_v'}{y^+}^4+O({y^+}^5)
                                                                                                                                    \notag\\
d_{\varepsilon_{zz}}^{(u)+}\sim&-4\overline{{A_w'^+}^2B_v'^+} {y^+}-6\left(\overline{{A_w'^+}^2C_v'^+}+4\overline{A_w'^+B_v'^+B_w'^+}\right){y^+}^2+O({y^+}^3)
                                                                                                                                    \notag
\end{alignat}
\vspace{-0.25in}
\begin{alignat}{6}
\Pi_{\varepsilon_{xx}}^+\sim& 8\overline{B_v'^+\dfrac{\partial A_u'^+}{\partial x^+}}
                           + 8\left( 3\overline{C_v'^+\dfrac{\partial A_u'^+}{\partial x^+}}
                                          + 2\overline{B_v'^+\dfrac{\partial B_u'^+}{\partial x^+}}
                                          - \overline{(\nabla A_u')^+\cdot(\nabla B_u')^+} \right ){y^+}
                             +O({y^+}^2)
                                                                                                                                    \notag\\
\Pi_{\varepsilon_{xy}}^+\sim&-12\overline{A_u'^+C_v'^+}
                             - 4\left(12\overline{A_u'^+D_v'^+}
                                     + 6\overline{B_v'^+C_u'^+}
                                     -  \overline{(\nabla A_u')^+\cdot(\nabla B_v')^+} \right ){y^+}
                             +O({y^+}^2)
                                                                                                                                    \notag\\
\Pi_{\varepsilon_{yy}}^+\sim& -    48\overline{B_v'^+C_v'^+}{y^+}
                               +8\left(-24\overline{B_v'^+D_v'^+}
                                      -  9\overline{{C_v'^+}^2}
                                      +   \overline{{(\nabla B_v')^+}^2} \right ){y^+}^2
                               +O({y^+}^3)
                                                                                                                                    \notag\\
\Pi_{\varepsilon_{zz}}^+\sim&       8\overline{B_v'^+\dfrac{\partial A_w'^+}{\partial z^+}}
                                 +8 \left(3\overline{C_v'^+\dfrac{\partial A_w'^+}{\partial z^+}}
                                         +2\overline{B_v'^+\dfrac{\partial B_w'^+}{\partial z^+}}
                                         - \overline{(\nabla A_w')^+\cdot(\nabla B_w')^+} \right ){y^+}
                           +O({y^+}^2)
                                                                                                                                    \notag
\end{alignat}
\vspace{-0.25in}
\begin{alignat}{6}
d_{\varepsilon_{xx}}^{(p)+}=&
d_{\varepsilon_{zz}}^{(p)+}=0\qquad\forall y^+
                                                                                                                                    \notag\\
d_{\varepsilon_{xy}}^{(p)+}\sim&- 8\left(\overline{B_u'^+B_v'^+}+3\overline{A_u'^+C_v'^+}\right)
                                -48\left(\overline{A_u'^+D_v'^+}+\overline{B_v'^+C_u'^+}\right){y^+}
                                +O({y^+}^2)
                                                                                                                                    \notag\\
d_{\varepsilon_{yy}}^{(p)+}\sim&- 16\overline{{B_v'^+}^2}
                                -192\overline{B_v'^+C_v'^+}{y^+}
                                -  8\left(48\overline{B_v'^+D_v'^+}
                                         +36\overline{{C_v'^+}^2}\right ){y^+}^2
                                +O({y^+}^3)
                                                                                                                                    \notag
\end{alignat}
\vspace{-0.25in}
\begin{alignat}{6}
\phi_{\varepsilon_{xx}}^+=&
\Pi_{\varepsilon_{xx}}^+\qquad\forall y^+
                                                                                                                                    \notag\\
\phi_{\varepsilon_{xy}}^+\sim&\left(8\overline{B_u'^+B_v'^+}+12\overline{A_u'^+C_v'^+}\right)
                             + 4\left( 6\overline{B_v'^+C_u'^+}
                                     +  \overline{(\nabla A_u')^+\cdot(\nabla B_v')^+} \right ){y^+}
                             +O({y^+}^2)
                                                                                                                                    \notag\\
\phi_{\varepsilon_{yy}}^+\sim&   16\overline{{B_v'^+}^2}
                               +144\overline{B_v'^+C_v'^+}{y^+}
                               +8\left( 24\overline{B_v'^+D_v'^+}
                                      + 27\overline{{C_v'^+}^2}
                                      +   \overline{{(\nabla B_v')^+}^2} \right ){y^+}^2
                               +O({y^+}^3)
                                                                                                                                    \notag\\
\phi_{\varepsilon_{zz}}^+=&
\Pi_{\varepsilon_{zz}}^+\qquad\forall y^+
                                                                                                                                    \notag
\end{alignat}
\vspace{-0.15in}
}

%% file: Tab_FluidDynRes_01_WAs_epsepsij_Pepsij.tex
\scalefont{.8}{
\begin{alignat}{6}
\varepsilon_{\varepsilon_{xx}}^+\sim&8\left(2\,\overline{{B_u'^+}^2}+\overline{{(\nabla A_u')^+}^2}\right )
                                 + 32\left(3\,\overline{B_u'^+\,C_u'^+}+\overline{(\nabla A_u')^+\cdot(\nabla B_u')^+}\right )y^++O({y^+}^2)
                                                                                                                                    \notag\\
\varepsilon_{\varepsilon_{xy}}^+\sim&16\overline{B_u'^+B_v'^+}  
                                    +16\left(3\overline{B_v'^+C_u'^+}+\overline{(\nabla A_u')^+\cdot(\nabla B_v')^+}\right ){y^+}+O({y^+}^2)
                                                                                                                                    \notag\\
\varepsilon_{\varepsilon_{yy}}^+\sim&16\,\overline{{B_v'^+}^2}+96\,\overline{B_v'^+C_v'^+}{y^+}+O({y^+}^2)
                                                                                                                                    \notag\\
\varepsilon_{\varepsilon_{zz}}^+\sim&8\left(2\,\overline{{B_w'^+}^2}+\overline{{(\nabla A_w')^+}^2}\right )
                                 + 32\left(3\overline{B_w'^+C_w'^+}+\overline{(\nabla A_w')^+\cdot(\nabla B_w')^+}\right )y^++O({y^+}^2)
                                                                                                                                    \notag
\end{alignat}
\vspace{-0.2in}
\begin{alignat}{6}
P_{\varepsilon_{xx}}^{(1)+}\sim&-8\,\overline{A_u'^+B_v'^+}{y^+}
                             + \left( \dfrac{8 \overline{A_u'^+B_v'^+}}{Re_{\tau_w}}
                                    -12\overline{A_u'^+C_v'^+}
                                    -16\overline{B_u'^+B_v'^+} \right)   {y^+}^2
                               +  O({y^+}^3)
                                                                                                                                    \notag\\
P_{\varepsilon_{xy}}^{(1)+}\sim&-8\,\overline{{B_v'^+}^2}{y^+}^2
                             +8 \left(  \dfrac{\overline{{B_v'^+}^2}}{Re_{\tau_w}}
                                     -3\,\overline{B_v'^+C_v'^+}                          \right ){y^+}^3
                               +  O({y^+}^4)
                                                                                                                                    \notag\\
P_{\varepsilon_{yy}}^{(1)+}   =&
P_{\varepsilon_{zz}}^{(1)+}   = 0\qquad\forall y^+
                                                                                                                                    \notag
\end{alignat}
\vspace{-0.2in}
\begin{alignat}{6}
P_{\varepsilon_{xx}}^{(2)+}\sim&-4\overline{B_u'^+\dfrac{\partial A_u'^+}{\partial x^+}}{y^+}^2-4\left( 2\overline{C_u'^+\dfrac{\partial A_u'^+}{\partial x^+}}-\dfrac{1}{Re_{\tau_w}}\overline{B_u'^+\dfrac{\partial A_u'^+}{\partial x^+}}\right){y^+}^3+O({y^+}^4)
                                                                                                                                    \notag\\
P_{\varepsilon_{xy}}^{(2)+}\sim&-2\overline{B_v'^+\dfrac{\partial  A_u'^+}{\partial x^+}}{y^+}^2-2\left( 2\overline{C_v'^+\dfrac{\partial A_u'^+}{\partial x^+}}-\dfrac{1}{Re_{\tau_w}}\overline{B_v'^+\dfrac{\partial A_u'^+}{\partial x^+}}\right){y^+}^3+O({y^+}^4)
                                                                                                                                    \notag\\
P_{\varepsilon_{yy}}^{(2)+}\sim&-4\,\overline{C_v'^+\dfrac{\partial B_v'^+}{\partial x^+}}{y^+}^4-4\left( 2\overline{D_v'^+\dfrac{\partial B_v'^+}{\partial x^+}}-\dfrac{1}{Re_{\tau_w}}\overline{C_v'^+\dfrac{\partial B_v'^+}{\partial x^+}}\right){y^+}^5+O({y^+}^6)
                                                                                                                                    \notag\\
P_{\varepsilon_{zz}}^{(2)+}\sim&-4\overline{B_w'^+\dfrac{\partial A_w'^+}{\partial x^+}}{y^+}^2-4\left( 2\overline{C_w'^+\dfrac{\partial A_w'^+}{\partial x^+}}-\dfrac{1}{Re_{\tau_w}}\overline{B_w'^+\dfrac{\partial A_w'^+}{\partial x^+}}\right){y^+}^3+O({y^+}^4)
                                                                                                                                    \notag
\end{alignat}
\vspace{-0.2in}
\begin{alignat}{6}
P_{\varepsilon_{xx}}^{(3)+}\sim& \dfrac{4}{Re_{\tau_w}}\,\overline{A_u'^+B_v'^+} {y^+}^2
                               + \dfrac{4}{Re_{\tau_w}}\left( \overline{A_u'^+C_v'^+}
                                                            +2\overline{B_u'^+B_v'^+}\right ){y^+}^3
                               +  O({y^+}^4)
                                                                                                                                    \notag\\
P_{\varepsilon_{xy}}^{(3)+}\sim& \dfrac{4}{Re_{\tau_w}}\,\overline{{B_v'^+}^2} {y^+}^3
                                +  \dfrac{10}{Re_{\tau_w}}\,\overline{B_v'^+C_v'^+} {y^+}^4
                               +  O({y^+}^5)
                                                                                                                                    \notag\\
P_{\varepsilon_{yy}}^{(3)+}   =&
P_{\varepsilon_{zz}}^{(3)+}   = 0\qquad\forall y^+
                                                                                                                                    \notag
\end{alignat}
\vspace{-0.2in}
\begin{alignat}{6}
P_{\varepsilon_{xx}}^{(4)+}\sim&-12\overline{{A_u'^+}^2B_v'^+} y^+
                            - 4\left( 6 \overline{{A_u'^+}^2C_v'^+}
                                    + 8 \overline{A_u'^+B_u'^+B_v'^+}
                                     -  \overline{A_u'^+B_u'^+\dfrac{\partial A_w'^+}{\partial z^+}}
                                     +  \overline{A_w'^+B_u'^+\dfrac{\partial A_u'^+}{\partial z^+}}\right){y^+}^2                  \notag\\
                              & +  O({y^+}^3)
                                                                                                                                    \notag\\
P_{\varepsilon_{xy}}^{(4)+} \sim& -2\left(  8\overline{A_u'^+{B_v'^+}^2}
                                          -  \overline{A_u'^+B_v'^+\dfrac{\partial A_w'^+}{\partial z^+}}
                                          +  \overline{A_w'^+B_v'^+\dfrac{\partial A_u'^+}{\partial z^+}} \right) {y^+}^2
                                                                                                                                    \notag\\
                              -4&\left(  16\overline{A_u'^+B_v'^+C_v'^+}
                                        + 8\overline{B_u'^+{B_v'^+}^2}
                                        -  \overline{B_u'^+B_v'^+\dfrac{\partial A_w'^+}{\partial z^+}}
                                        +  \overline{B_w'^+B_v'^+\dfrac{\partial A_u'^+}{\partial z^+}}
                                        +  \overline{A_w'^+C_v'^+\dfrac{\partial A_u'^+}{\partial z^+}}
                                        +  \overline{A_u'^+C_v'^+\dfrac{\partial A_u'^+}{\partial x^+}}\right) {y^+}^3             \notag\\
                               &+  O({y^+}^4)
                                                                                                                                    \notag\\
P_{\varepsilon_{yy}}^{(4)+}\sim&-40\overline{{B_v'^+}^3}{y^+}^3
                                 -4 \left( 46\overline{{B_v'^+}^2C_v'^+}
                                          +  \overline{A_u'^+C_v'^+\dfrac{\partial B_v'^+}{\partial x^+}}
                                          +  \overline{A_w'^+C_v'^+\dfrac{\partial B_v'^+}{\partial z^+}} \right) {y^+}^4
                                        +  O({y^+}^5)
                                                                                                                                    \notag\\
P_{\varepsilon_{zz}}^{(4)+}\sim&-12\overline{{A_w'^+}^2B_v'^+} y^+
                              -4\left( 6\overline{{A_w'^+}^2C_v'^+}
                                     + 8\overline{A_w'^+B_w'^+B_v'^+}
                                     -  \overline{A_w'^+B_w'^+\dfrac{\partial A_u'^+}{\partial x^+}}
                                     +  \overline{A_u'^+B_w'^+\dfrac{\partial A_w'^+}{\partial x^+}}\right){y^+}^2                  \notag\\
                               &+  O({y^+}^3)
                                                                                                                                    \notag
\end{alignat}
\vspace{-0.15in}
}

%% file: Tab_FluidDynRes_01_WAs_epseps.tex
\scalefont{1.0}{
\begin{alignat}{6}
\varepsilon_{\varepsilon_{xx}}^+\sim& 8\left(2\,\overline{{B_u'^+}^2}+\overline{{(\nabla A_u')^+}^2}\right)
                                   + 32\left(3\,\overline{B_u'^+\,C_u'^+}+\overline{(\nabla A_u')^+\cdot(\nabla B_u')^+}\right )y^++O({y^+}^2)
                                                                                                                                    \notag\\
\varepsilon_{\varepsilon_{xy}}^+\sim&16\,\overline{B_u'^+B_v'^+}  
                                   + 16\left(3\overline{B_v'^+C_u'^+}+{\color{blue}{\left\{3\overline{B_u'^+C_v'^+}\right\}}}+\overline{(\nabla A_u')^+\cdot(\nabla B_v')^+}\right){y^+}+O({y^+}^2)
                                                                                                                                    \notag\\
\varepsilon_{\varepsilon_{yy}}^+\sim&16\,\overline{{B_v'^+}^2}+96\,\overline{B_v'^+C_v'^+}{y^+}+O({y^+}^2)
                                                                                                                                    \notag\\
\varepsilon_{\varepsilon_{yz}}^+\sim&{\color{blue}{\Bigg[
                                     16\,\overline{B_w'^+B_v'^+}  
                                   + 16\left(3\overline{B_v'^+C_w'^+}+3\overline{B_w'^+C_v'^+}+\overline{(\nabla A_w')^+\cdot(\nabla B_v')^+}\right){y^+}+O({y^+}^2)
                                    \Bigg]}}
                                                                                                                                    \notag\\
\varepsilon_{\varepsilon_{zz}}^+\sim& 8\left(2\,\overline{{B_w'^+}^2}+\overline{{(\nabla A_w')^+}^2}\right)
                                   + 32\left(3\overline{B_w'^+C_w'^+}+\overline{(\nabla A_w')^+\cdot(\nabla B_w')^+}\right)y^++O({y^+}^2)
                                                                                                                                    \notag\\
\varepsilon_{\varepsilon_{zx}}^+\sim&{\color{blue}{\Bigg[
                                      8\left(2\,\overline{B_u'^+B_w'^+}+\overline{(\nabla A_u')^+\cdot(\nabla A_w')^+}\right)
                                     }}
                                                                                                                                    \notag\\
                                     {\color{blue}{
                                   + 16}}
                                     &{\color{blue}{
                                      \left(3\overline{B_u'^+C_w'^+}+3\overline{B_w'^+C_u'^+}+\overline{(\nabla A_u')^+\cdot(\nabla B_w')^+}+\overline{(\nabla A_w')^+\cdot(\nabla B_u')^+}\right){y^+}+O({y^+}^2)
                                      \Bigg]}}
                                                                                                                                    \notag
\end{alignat}
}

%% file: Tab_FluidDynRes_01_WAs_bepseps.tex
\scalefont{.65}{
\begin{alignat}{6}
     b_{\epsilon_{\epsilon_{xx}}}    \sim&-\tfrac{1}
                                                 {3}\dfrac{2\overline{{B_w'^+}^2}+2\overline{{B_v'^+}^2}-4\overline{{B_u'^+}^2}+\overline{{(\nabla A_w')^+}^2}-2\overline{{(\nabla A_u')^+}^2}}
                                                          {2\overline{{B_u'^+}^2}+2\overline{{B_v'^+}^2}+2\overline{{B_w'^+}^2}+\overline{{(\nabla A_u')^+}^2}+\overline{{(\nabla A_w')^+}^2} }
                                                                                                                                    \notag\\
                                         &+4\dfrac{-\left(2\overline{{B_u'^+}^2}+\overline{{(\nabla A_u')^+}^2}                       \right)\left(3\overline{B_w'^+C_w'^+}+3\overline{B_v'^+C_v'^+}+\overline{(\nabla A_w')^+\cdot(\nabla B_w')^+}\right)
                                                   +\left(2\overline{{B_w'^+}^2}+2\overline{{B_v'^+}^2}+\overline{{(\nabla A_w')^+}^2}\right)\left(3\overline{B_u'^+C_u'^+}+\overline{(\nabla A_u')^+\cdot(\nabla B_u')^+}\right)}
                                                  {\left(2\overline{{B_u'^+}^2}+2\overline{{B_v'^+}^2}+2\overline{{B_w'^+}^2}+\overline{{(\nabla A_u')^+}^2}+\overline{{(\nabla A_w')^+}^2}\right)^2}{y^+}+O({y^+}^2)
                                                                                                                                    \notag\\
     b_{\epsilon_{\epsilon_{xy}}}    \sim&          \dfrac{2\overline{B_u'^+B_v'^+}                                                                                                          }
                                                          {2\overline{{B_u'^+}^2}+2\overline{{B_v'^+}^2}+2\overline{{B_w'^+}^2}+\overline{{(\nabla A_u')^+}^2}+\overline{{(\nabla A_w')^+}^2}}
                                          +2\Vast(\dfrac{ \left(2\overline{{B_u'^+}^2}+2\overline{{B_v'^+}^2}+2\overline{{B_w'^+}^2}+\overline{{(\nabla A_u')^+}^2}+\overline{{(\nabla A_w')^+}^2}\right)
                                                          \left(3\overline{B_v'^+C_u'^+}+{\color{blue}{\left\{3\overline{B_u'^+C_v'^+}\right\}}}+\overline{(\nabla A_u')^+\cdot(\nabla B_v')^+}\right)}
                                                        {\left(2\overline{{B_u'^+}^2}+2\overline{{B_v'^+}^2}+2\overline{{B_w'^+}^2}+\overline{{(\nabla A_u')^+}^2}+\overline{{(\nabla A_w')^+}^2}\right)^2}
                                                                                                                                    \notag\\
                                         &       -\dfrac{4\overline{B_u'^+B_v'^+}\left(3\overline{B_u'^+C_u'^+}+3\overline{B_v'^+C_v'^+}+3\overline{B_w'^+C_w'^+}+\overline{(\nabla A_u')^+\cdot(\nabla B_u')^+}+\overline{(\nabla A_w')^+\cdot(\nabla B_w')^+}\right)}
                                                        {\left(2\overline{{B_u'^+}^2}+2\overline{{B_v'^+}^2}+2\overline{{B_w'^+}^2}+\overline{{(\nabla A_u')^+}^2}+\overline{{(\nabla A_w')^+}^2}\right)^2}\vast){y^+}+O({y^+}^2)
                                                                                                                                    \notag\\
     b_{\epsilon_{\epsilon_{yy}}}    \sim&-\tfrac{1}
                                                 {3}\dfrac{2\overline{{B_u'^+}^2}-4\overline{{B_v'^+}^2}+2\overline{{B_w'^+}^2}+\overline{{(\nabla A_u')^+}^2}+\overline{{(\nabla A_w')^+}^2}}
                                                          {2\overline{{B_u'^+}^2}+2\overline{{B_v'^+}^2}+2\overline{{B_w'^+}^2}+\overline{{(\nabla A_u')^+}^2}+\overline{{(\nabla A_w')^+}^2} }
                                                                                                                                    \notag\\
                                         &+4\dfrac{3\overline{B_v'^+C_v'^+}\left(2\overline{{B_u'^+}^2}+2\overline{{B_w'^+}^2}+\overline{{(\nabla A_u')^+}^2}+\overline{{(\nabla A_w')^+}^2}\right)
                                                   -2\overline{{B_v'^+}^2}\left(3\overline{B_u'^+C_u'^+}+3\overline{B_w'^+C_w'^+}+\overline{(\nabla A_u')^+\cdot(\nabla B_u')^+}+\overline{(\nabla A_w')^+\cdot(\nabla B_w')^+}\right)}
                                                  {\left(2\overline{{B_u'^+}^2}+2\overline{{B_v'^+}^2}+2\overline{{B_w'^+}^2}+\overline{{(\nabla A_u')^+}^2}+\overline{{(\nabla A_w')^+}^2}\right)^2}{y^+}+O({y^+}^2)
                                                                                                                                    \notag\\
     b_{\epsilon_{\epsilon_{yz}}}     \sim&{\color{Blue}{\Vast[
                                                    \dfrac{2\overline{B_w'^+B_v'^+}                                                                                                          }
                                                          {2\overline{{B_u'^+}^2}+2\overline{{B_v'^+}^2}+2\overline{{B_w'^+}^2}+\overline{{(\nabla A_u')^+}^2}+\overline{{(\nabla A_w')^+}^2}}
                                          +2\Vast(\dfrac{ \left(2\overline{{B_u'^+}^2}+2\overline{{B_v'^+}^2}+2\overline{{B_w'^+}^2}+\overline{{(\nabla A_u')^+}^2}+\overline{{(\nabla A_w')^+}^2}\right)
                                                          \left(3\overline{B_v'^+C_w'^+}+3\overline{B_w'^+C_v'^+}+\overline{(\nabla A_w')^+\cdot(\nabla B_v')^+}\right)}
                                                        {\left(2\overline{{B_u'^+}^2}+2\overline{{B_v'^+}^2}+2\overline{{B_w'^+}^2}+\overline{{(\nabla A_u')^+}^2}+\overline{{(\nabla A_w')^+}^2}\right)^2}
                                           }}
                                                                                                                                    \notag\\
                                         &{\color{Blue}{
                                                 -\dfrac{4\overline{B_v'^+B_w'^+}\left(3\overline{B_u'^+C_u'^+}+3\overline{B_v'^+C_v'^+}+3\overline{B_w'^+C_w'^+}+\overline{(\nabla A_u')^+\cdot(\nabla B_u')^+}+\overline{(\nabla A_w')^+\cdot(\nabla B_w')^+}\right)}
                                                        {\left(2\overline{{B_u'^+}^2}+2\overline{{B_v'^+}^2}+2\overline{{B_w'^+}^2}+\overline{{(\nabla A_u')^+}^2}+\overline{{(\nabla A_w')^+}^2}\right)^2}\vast){y^+}+O({y^+}^2)
                                           \Vast]}}
                                                                                                                                    \notag\\
     b_{\epsilon_{\epsilon_{zz}}}    \sim&-\tfrac{1}
                                                 {3}\dfrac{2\overline{{B_u'^+}^2}+2\overline{{B_v'^+}^2}-4\overline{{B_w'^+}^2}+\overline{{(\nabla A_u')^+}^2}-2\overline{{(\nabla A_w')^+}^2}}
                                                          {2\overline{{B_u'^+}^2}+2\overline{{B_v'^+}^2}+2\overline{{B_w'^+}^2}+\overline{{(\nabla A_u')^+}^2}+\overline{{(\nabla A_w')^+}^2} }
                                                                                                                                    \notag\\
                                         &+4\dfrac{-\left(2\overline{{B_w'^+}^2}+\overline{{(\nabla A_w')^+}^2}                       \right)\left(3\overline{B_u'^+C_u'^+}+3\overline{B_v'^+C_v'^+}+\overline{(\nabla A_u')^+\cdot(\nabla B_u')^+}\right)
                                                   +\left(2\overline{{B_u'^+}^2}+2\overline{{B_v'^+}^2}+\overline{{(\nabla A_u')^+}^2}\right)\left(3\overline{B_w'^+C_w'^+}+\overline{(\nabla A_w')^+\cdot(\nabla B_w')^+}\right)}
                                                  {\left(2\overline{{B_u'^+}^2}+2\overline{{B_v'^+}^2}+2\overline{{B_w'^+}^2}+\overline{{(\nabla A_u')^+}^2}+\overline{{(\nabla A_w')^+}^2}\right)^2}{y^+}+O({y^+}^2)
                                                                                                                                    \notag\\
     b_{\epsilon_{\epsilon_{zx}}}     \sim&{\color{Blue}{\Vast[
                                                    \dfrac{2\overline{B_u'^+B_w'^+}+\overline{(\nabla A_u')^+\cdot(\nabla A_w')^+}                                                           }
                                                          {2\overline{{B_u'^+}^2}+2\overline{{B_v'^+}^2}+2\overline{{B_w'^+}^2}+\overline{{(\nabla A_u')^+}^2}+\overline{{(\nabla A_w')^+}^2}}
                                          +2\Vast(\dfrac{ \left(2\overline{{B_u'^+}^2}+2\overline{{B_v'^+}^2}+2\overline{{B_w'^+}^2}+\overline{{(\nabla A_u')^+}^2}+\overline{{(\nabla A_w')^+}^2}\right)
                                                          \left(3\overline{B_u'^+C_w'^+}+3\overline{B_w'^+C_u'^+}+\overline{(\nabla A_u')^+\cdot(\nabla B_w')^+}+\overline{(\nabla A_w')^+\cdot(\nabla B_u')^+}\right)}
                                                        {\left(2\overline{{B_u'^+}^2}+2\overline{{B_v'^+}^2}+2\overline{{B_w'^+}^2}+\overline{{(\nabla A_u')^+}^2}+\overline{{(\nabla A_w')^+}^2}\right)^2}
                                           }}
                                                                                                                                    \notag\\
                                         &{\color{Blue}{
                                                 -\dfrac{2\left(2\overline{B_u'^+B_w'^+}+\overline{(\nabla A_u')^+\cdot(\nabla A_w')^+}\right)
                                                          \left(3\overline{B_u'^+C_u'^+}+3\overline{B_v'^+C_v'^+}+3\overline{B_w'^+C_w'^+}+\overline{(\nabla A_u')^+\cdot(\nabla B_u')^+}+\overline{(\nabla A_w')^+\cdot(\nabla B_w')^+}\right)}
                                                        {\left(2\overline{{B_u'^+}^2}+2\overline{{B_v'^+}^2}+2\overline{{B_w'^+}^2}+\overline{{(\nabla A_u')^+}^2}+\overline{{(\nabla A_w')^+}^2}\right)^2}\vast){y^+}+O({y^+}^2)
                                           \Vast]}}
                                                                                                                                    \notag
\end{alignat}
}